\documentclass[twocolumn,dvipsnames]{aastex701}
\usepackage{amsmath,amstext}
\usepackage[figure,figure*]{hypcap}
\usepackage{ulem}
\usepackage{xspace}
\usepackage{xcolor}
\usepackage{lineno}
\usepackage{xfrac}
\usepackage{graphicx}
\usepackage{multimedia}
\usepackage{amssymb}
\usepackage{amsmath}
\usepackage{array,booktabs}
\usepackage{mathptmx}
\usepackage{booktabs}
\usepackage{hyperref}
\usepackage{float}
\usepackage{todonotes}
\usepackage{multirow}
\newcommand{\s}{\,{\rm s}}

\newcommand{\g}{\,{\rm g}}

\newcommand{\mbh}{\,{M_{\rm BH}}}

\graphicspath{{./}{figures/}}

\shorttitle{Accretion of PBHs in Stars}
\shortauthors{Cantiello, Gottlieb, Norton, Kleban \& Van Tilburg}

\begin{document}
\title{
%Radiative Efficiency and Growth of Primordial Black Holes in Stellar Interiors
Accretion of Primordial Black Holes in Stellar Interiors
}
\correspondingauthor{Matteo Cantiello}
\email{mcantiello@flatironinstitute.org}

\author[0000-0002-8171-8596]{Matteo Cantiello} 
\affil{Center for Computational Astrophysics, Flatiron Institute, 162 5th Avenue, New York, NY 10010, USA} 
\affil{Department of Astrophysical Sciences, Princeton University, Princeton, NJ 08544, USA}
\email{mcantiello@flatironinstitute.org}
\author[0000-0003-3115-2456]{Ore Gottlieb}
\affil{Department of Physics and Kavli Institute for Astrophysics and Space Research, Massachusetts Institute of Technology, Cambridge, MA 02139, USA}
\email{oregottlieb@gmail.com }
\author[0000-0002-4382-9885]{Cameron Norton}
\affil{Center for Cosmology and Particle Physics, Department of Physics, New York University, New York, NY 10003 USA}
\email{cen8858@nyu.edu}
\author[0000-0002-1889-2487]{Matthew Kleban}
\affil{Center for Cosmology and Particle Physics, Department of Physics, New York University, New York, NY 10003 USA}
\email{kleban@nyu.edu}
\author[0000-0001-7085-6128]{Ken Van Tilburg}
\email{kenvt@stanford.edu}
\affil{Leinweber Institute for Theoretical Physics, Department of Physics, Stanford University, Stanford, CA 94305, USA}

\begin{abstract}
We study spherical accretion onto primordial black holes (PBHs) embedded in the core of a solar-type star. We compute the radiative efficiency self-consistently for the first time across the optically thin range ($10^{-16.5}$--$10^{-10}\,M_\odot$) with time-dependent simulations, and follow the growth up to $10^{-2}\,M_\odot$ using an analytical photon-trapping prescription above $5\times 10^{-13}\,M_\odot$. Near the Schwarzschild radius ($r_{\rm S} \sim 10^{-11}$\,cm for a $10^{-16}\,M_\odot$ PBH), gas compressed to temperatures up to $\sim 10^{11}$\,K radiates through microphysical processes that fundamentally alter the classical adiabatic Bondi solution. We solve the time-dependent spherical Euler equations with an implicit cooling source term, self-consistently determining the accretion rate $\dot{M}$, the radiative efficiency $\eta = L/\dot{M}c^2$, and the flow structure. We identify three physically distinct regimes for spherical accretion. At low masses ($M_{\rm BH} \lesssim 10^{-14}\,M_\odot$) the flow is in a ``Hot Bondi'' state: bremsstrahlung cooling is dynamically negligible, the gas reaches $T\sim 10^{11}$\,K near $r_{\rm S}$. In the bremsstrahlung cooling regime ($10^{-14}$--$5\times 10^{-13}\,M_\odot$), increasingly efficient cooling drives the flow toward isothermal, with $\eta \approx 10^{-2}$. Above $\sim 5\times 10^{-13}\,M_\odot$ the Bondi sphere becomes optically thick, entering a photon-trapping regime with the accretion rate remaining close to the Bondi value.
%We include radiative feedback on the boundary conditions through both photon diffusion and convective (MLT) envelope models, and find that feedback is negligible in the Hot Bondi regime but moderates accretion at higher masses.
Cooling enhances the accretion rate by a factor of $\sim$2--7 relative to the adiabatic Bondi rate, substantially increasing the black hole growth rate, which stays super-exponential throughout the spherical regime. The radiative efficiency is an order of magnitude lower than previously assumed, and the critical initial PBH mass required to consume a solar-mass star within a Hubble time is $M_{\rm 0,crit} \sim 10^{-16}\,M_\odot$.
\end{abstract}

\section{Introduction}
\label{sec:introduction}

Primordial black holes (PBHs)---black holes formed 
in the early Universe \citep{Zeldovich1967,Hawking1971,Carr1974}---remain a
viable dark matter candidate across a wide range of masses
\citep{Chapline1975,Carr2016,Ali_Ha_moud_2017, Green2021,Villanueva-Domingo2021, Escriva2024, Garriga_2016, Kleban:2023ugf, Crawford:1982yz, Rubin:2000dq}.
Observational constraints from microlensing, gravitational waves, and
cosmological probes have excluded PBHs as the dominant dark matter component
in several mass windows, but the sub-lunar range
($10^{-16} \, M_\odot \lesssim M_\mathrm{BH} \lesssim 10^{-10}\,M_\odot$) remains poorly constrained
\citep{Montero-Camacho2019,Green2021}.

Stars provide a unique laboratory for probing PBHs in this mass range. A PBH captured by (or formed within) a stellar interior will accrete the surrounding gas and release gravitational energy, potentially altering the star's structure and evolution \citep{Hawking1971}. \citet{Bellinger2023} recently explored this scenario using 1D stellar evolution models computed with the \texttt{MESA} code \citep{Paxton2011}, finding that PBHs
with initial masses $M_0 \gtrsim 10^{-11}\,M_\odot$ can significantly
affect the evolution of solar-type stars and may be constrained by
asteroseismic observations.  Their treatment adopts a fixed radiative
efficiency $\eta = 0.08$, the canonical value for a geometrically
thin accretion disk extending to the innermost stable circular orbit of a
Schwarzschild black hole, and applies the Eddington luminosity limit at
all masses.

However, the accretion geometry for a PBH embedded in a stellar core
differs fundamentally from the thin-disk paradigm.  For $M_{\rm BH}\lesssim 10^{-10}\,M_\odot$, the Bondi radius
$r_\mathrm{B} = G M_\mathrm{BH}/c_\infty^2$ ranges from
$\sim$$500$\,\AA\ ($M_\mathrm{BH} = 10^{-16}\,M_\odot$) to
$\sim$$5$\,cm ($10^{-10}\,M_\odot$)---far smaller than the stellar
circularization radius, so no accretion disk forms.  Instead, gas accretes
spherically, passing through a sonic point at $\sim r_\mathrm{B}$ and
accelerating to relativistic speeds near the Schwarzschild radius
$r_\mathrm{S} = 2GM_\mathrm{BH}/c^2$, which lies five orders of magnitude
inward.  In the process, the gas is compressed and heated to temperatures
$T \sim 10^{10}$--$10^{12}$\,K, where relativistic bremsstrahlung, pair
processes, and Klein--Nishina scattering effects dominate the
microphysics.  The appropriate radiative efficiency must be calculated
self-consistently from these processes, not assumed from thin-disk theory.

Moreover, the Eddington limit depends on the optical depth of the Bondi sphere.  For
$M_\mathrm{BH} \lesssim 10^{-14}\,M_\odot$, the cumulative optical depth
through the accretion flow (including Klein--Nishina suppression of the
scattering cross-section and the density enhancement within $r_\mathrm{B}$)
is below unity, and photons produced near $r_\mathrm{S}$ escape the
accretion region without significant interaction.  In this optically thin
regime, the Eddington limit is irrelevant regardless of the luminosity.
At higher masses, where the Bondi sphere becomes optically thick, the
luminosity must be self-consistently compared to the Eddington value using
the appropriate (Klein--Nishina--corrected) opacity.

In this paper, we compute the accretion rate, radiative efficiency
$\eta = L/\dot{M}c^2$, and flow structure for a PBH accreting from a
solar-type stellar core, spanning more than six decades in black hole mass
($10^{-16.5}$--$10^{-10}\,M_\odot$).  We solve the time-dependent
spherical Euler equations with an implicit bremsstrahlung and pair-process
cooling source term, evolving the flow from an initial adiabatic Bondi
profile to a self-consistent steady state.  The Euler solver yields
the physical $\eta$ in the optically thin range
($M_\mathrm{BH}\lesssim 5\times 10^{-13}\,M_\odot$);
at higher masses the Bondi sphere becomes optically thick and we treat
photon trapping with the analytical prescription of \citet{Begelman1979}.  This approach avoids the
singular-point difficulties inherent in steady-state ODE methods, which
fail when cooling is dynamically important (the sonic point becomes a
focus rather than a saddle).

The paper is organized as follows.  Section~\ref{sec:methods} describes
the modified Bondi accretion problem, the numerical method, opacity
considerations, and the radiative feedback model.
Section~\ref{sec:results} presents the accretion regimes, the
self-consistently computed radiative efficiency, and the resulting PBH
growth trajectories.  Section~\ref{sec:discussion} discusses magnetic
flux accumulation, comparison with previous work, and relation to other
radiating Bondi flow calculations.  Section~\ref{sec:conclusions}
summarizes our conclusions.

\section{Radiating Bondi Flows}
\label{sec:methods}

We consider a PBH of mass $M_\mathrm{BH}$ embedded
at the center of a solar-type star and derive the accretion rate, radiative
efficiency, and resulting growth trajectory from first principles.  We adopt
conditions appropriate for the solar core: temperature
$T_\infty = 1.57 \times 10^7$\,K, density
$\rho_\infty = 150$\,g\,cm$^{-3}$, hydrogen mass fraction $X = 0.34$,
helium mass fraction $Y = 0.64$, mean molecular weight $\mu = 0.85$, and
adiabatic index $\gamma = 5/3$.  The corresponding sound speed is
$c_\infty = \sqrt{\gamma k_\mathrm{B} T_\infty / \mu m_\mathrm{p}}
\approx 500$\,km\,s$^{-1}$.

\subsection{Microphysical Emissivity}
\label{sec:modified_bondi}

The classical Bondi accretion rate onto a point mass at rest in a uniform
medium is \citep{1952MNRAS.112..195B}
\begin{equation}
  \dot{M}_\mathrm{B} = 4\pi\,\lambda\,\rho_\infty\,
  \frac{(G M_\mathrm{BH})^2}{c_\infty^3}\,,
  \label{eq:Mdot_bondi}
\end{equation}
where $\lambda$ is a dimensionless eigenvalue set by the requirement that the
flow passes smoothly through the sonic point.
Its value depends on the adiabatic index: $\lambda = 1/4$ for $\gamma = 5/3$.
The relativistic generalization of this transonic problem onto a Schwarzschild
black hole was given by \citet{Michel:1972}; in the present application the
sonic point lies at $r\!\gg\!r_\mathrm{S}$, so the
Newtonian treatment is adequate.
The Bondi radius
$r_\mathrm{B} = G M_\mathrm{BH} / c_\infty^2$ delineates the sphere of
gravitational influence; for $M_\mathrm{BH} = 10^{-16}\,M_\odot$, roughly
the mass of a small asteroid,
$r_\mathrm{B} \approx 5 \times 10^{-6}$\,cm $\approx 500$\,\AA.  The
Schwarzschild radius
$r_\mathrm{S} = 2GM_\mathrm{BH}/c^2 \approx 3 \times 10^{-11}$\,cm lies
five orders of magnitude inward.  The ratio
$r_\mathrm{B}/r_\mathrm{S} = c^2/(2c_\infty^2) \approx 1.8 \times 10^5$
is independent of $M_\mathrm{BH}$ (at fixed ambient conditions), providing
a large and universal dynamic range over which the infalling gas is compressed
and heated.

In the adiabatic Bondi solution, the radial profiles of density, temperature,
and velocity follow power-law scalings at $r \ll r_\mathrm{B}$:
$\rho \propto r^{-3/2}$, $T \propto r^{-1}$, and
$v \propto r^{-1/2}$ (see dashed lines in Fig.~\ref{fig:profiles}).  The temperature rises from $T_\infty$ at large radii
to $T(r_\mathrm{S}) \sim T_\infty\,(r_\mathrm{B}/r_\mathrm{S}) \sim
10^{12}$\,K near the horizon, reaching the regime where the dimensionless
electron temperature $\theta_e \equiv k_\mathrm{B}T / m_e c^2 \gg 1$ and relativistic processes dominate the emissivity.

We modify the standard Bondi equations by adding a local volumetric cooling term
$\varepsilon(\rho, T)$ computed from the following microphysical processes:

\begin{enumerate}
\item \textbf{Bremsstrahlung (free--free emission).}  For
$\theta_e \ll 1$, we use the standard non-relativistic emissivity
$\varepsilon_\mathrm{ff} \propto n_e n_i T^{1/2}$,
where $n_e$ and $n_i$ are the electron and ion number densities
\citep{Rybicki1979}.  For $\theta_e \gtrsim 1$, we adopt the
relativistic fitting functions of \citet{Stepney1983}, which include
both electron--ion and electron--electron contributions.

\item \textbf{Electron--positron pair annihilation
($e^+e^- \to 2\gamma$).} At temperatures $T \gtrsim 6 \times 10^9$\,K ($\theta_e \gtrsim 1$), the thermal photon bath maintains an equilibrium population of $e^+e^-$ pairs whose annihilation provides an additional, steeply temperature-dependent source of radiation \citep{Svensson1982,Stepney1983}.

\end{enumerate}

Note that pair annihilation into neutrinos ($e^+e^- \to \nu\bar\nu$) proceeds via the weak interaction rather than electromagnetism, suppressing its cross-section by $\sim 10^{-20}$ relative to the photon channel at the temperatures reached in our calculations. For this reason we neglect it. At low BH masses, the Coulomb mean free path $\lambda_\mathrm{C} \propto T^2 / (n \ln\Lambda)$, where $n = \rho / \mu m_\mathrm{p}$ is the particle number density, grows steeply inward (the square of $T \propto r^{-1}$ rises faster than $n \propto r^{-3/2}$) and exceeds the local flow scale before the gas reaches pair-producing temperatures ($\theta_e \sim 1$). In this Hot Bondi regime, where the inner flow becomes collisionless before pair-producing temperatures are reached, bremsstrahlung (which requires only binary collisions) dominates the cooling (Section~\ref{sec:regimes}).

\subsection{Validity of the fluid description}
\label{sec:fluid_validity}

The Euler equations~\eqref{eq:euler} apply only where the gas is
collisional on the relevant flow scales.  In the solar core, the
electron--ion Coulomb mean free path at ambient conditions is
$\lambda_\mathrm{C}\sim 10^{-7}$--$10^{-8}$\,cm.  The Knudsen number
at the Bondi radius,
$\mathrm{Kn}(r_\mathrm{B}) \equiv \lambda_\mathrm{C}/r_\mathrm{B}
\propto M_\mathrm{BH}^{-1}$, is therefore largest at the lightest mass
in our range and equals $\mathrm{Kn}(r_\mathrm{B})\sim 10^{-2}$ at
$M_\mathrm{BH} = 10^{-16.5}\,M_\odot$.  The gas is thus solidly
collisional in the Bondi-rate-setting subsonic zone ($r\gtrsim
r_\mathrm{B}/4$ for the adiabatic flow, and farther out once cooling
moves the sonic point outward), so the fluid description is secure
where $\dot{M}$ is determined.

Inside the Bondi sphere, the adiabatic profiles give
$\lambda_\mathrm{C}\propto r^{-1/2}$, so $\lambda_\mathrm{C}/r\propto
r^{-3/2}$ and the gas becomes collisionless at a transition radius
$r_\mathrm{coll}$ well inside $r_\mathrm{B}$ for the lightest masses
(for $M_\mathrm{BH}=10^{-16}\,M_\odot$, $r_\mathrm{coll}\simeq 7000\,
r_\mathrm{S}\simeq 0.04\,r_\mathrm{B}$, still subsonic).  Within
$r_\mathrm{coll}$, plasma microinstabilities in the counter-streaming
inflow are expected to provide anomalous collisionality on the
ion-skin-depth scale; we discuss the corresponding loss-cone and
recycling corrections in Appendix~\ref{sec:collisionless_pileup} and
find that the net suppression of $\dot{M}$ is at most $\sim 30\%$ at
$10^{-16}\,M_\odot$, shrinking rapidly with mass.

The opposite extreme ---  a fully collisionless gas, in which
$r_\mathrm{B}$ itself becomes smaller than $\lambda_\mathrm{C}$ ---
suppresses the Bondi rate by a factor $\sim r_\mathrm{S}/r_\mathrm{B} =
2c_\infty^2/c^2$ relative to the hydrodynamic value, since the
gravitational capture cross-section drops from $\sim\pi r_\mathrm{B}^2$
(focused, fluid-supported) to $\sim\pi r_\mathrm{S}^2$ (single-particle
loss-cone) \citep{Eddington:1926,Danby:1957}.  The interpolation between
these limits requires solving the Vlasov equation for the particle
distribution function \citep{Begelman:1977}.  For solar-core conditions
this fully collisionless regime sets in only at
$M_\mathrm{BH}\lesssim c_\infty^2\,\lambda_\mathrm{C}/G \sim
10^{-18}\,M_\odot$, more than two decades below our range.  Throughout
the masses considered in this work, the supply of gas at $r_\mathrm{B}$
is therefore set by collisional Bondi accretion, and any kinetic
corrections are confined to the inner zone treated in
Appendix~\ref{sec:collisionless_pileup}.

We neglect any potential quantum-mechanical suppression of accretion on very small black holes (cf.~\cite{Loeb_2024}).  Taking Eq.~(7) of \cite{Loeb_2024} at face value,  for $M \gtrsim 10^{17}g$ the quantum-limited rate exceeds Bondi and is therefore irrelevant.

\subsection{Numerics}
\label{sec:sec_numerics}

The coupled ordinary differential equations (ODEs) governing $T(r)$ and $v(r)$ in
the presence of cooling contain a singular point where the flow velocity equals
the local sound speed ($v = c_s$).  In the adiabatic limit, this sonic point is
a saddle and the transonic solution can be obtained by standard shooting methods.
However, when cooling is dynamically important, i.e.~when the cooling time $t_\mathrm{cool}$ is shorter than the free-fall time $t_\mathrm{ff}$ (as it is for
$M_\mathrm{BH} \gtrsim 10^{-14}\,M_\odot$),  %, where $t_\mathrm{cool} \lesssim t_\mathrm{ff}$) 
the character of the singular point changes.  The eigenvalues of the linearized system at the sonic point become
complex (a focus rather than a saddle), and no smooth transonic solution
exists as a trajectory through the singular point.  This mathematical situation
reflects the physical fact that in the strongly cooled regime, the steady-state
flow structure cannot be captured by an ODE boundary value problem.

We therefore solve the full time-dependent spherical Euler equations
with an implicit cooling source term, evolving the flow from an initial
adiabatic Bondi profile to a self-consistent steady state. The
numerical implementation is provided by \texttt{radbondi},\footnote{%
\url{https://github.com/matteocantiello/radbondi}} an open-source
Python package developed for this work. The conservative form of the
equations is
\begin{align}
  \frac{\partial}{\partial t}
  \begin{pmatrix} \rho \\ \rho v \\ e \end{pmatrix}
  + \frac{1}{r^2} \frac{\partial}{\partial r} r^2
  \begin{pmatrix} \rho v \\ \rho v^2 + P \\ (e+P)v \end{pmatrix}
  &= \begin{pmatrix} 0 \\ -\rho\,GM/r^2 \\ -\rho v\,GM/r^2 - \varepsilon \end{pmatrix}\,,
  \label{eq:euler}
\end{align}
where $e = \frac{1}{2}\rho v^2 + P/(\gamma - 1)$ is the total energy density
and $\varepsilon(\rho, T)$ is the microphysical emissivity described above.
We solve these equations on a logarithmic grid in $r/r_\mathrm{B}$, spanning
from just outside $r_\mathrm{S}$ to $r_\mathrm{B}$, using the HLL approximate
Riemann solver \citep{HLL1983} with MUSCL-Hancock reconstruction for
second-order spatial accuracy.  Gravity source terms are treated with a
well-balanced scheme that preserves hydrostatic equilibrium to machine precision,
ensuring that the adiabatic Bondi solution is maintained exactly in the absence
of cooling.  The cooling term $\varepsilon$ is integrated implicitly at each
timestep via operator splitting, which is essential because the cooling time in
the inner cells can be orders of magnitude shorter than the CFL-limited
hydrodynamic timestep.

The outer boundary is held at the ambient stellar conditions
($\rho_\infty$, $T_\infty$, $v = 0$); the inner boundary is placed at
$r_\mathrm{in} \approx 0.5$--$2\,r_\mathrm{S}$ (depending on mass; see Table~\ref{tab:simulations}) and is supersonic
(all characteristics point inward), so no boundary condition is imposed---the
flow variables are extrapolated from the interior.  We adopt Newtonian gravity
throughout; general relativistic corrections modify the flow structure only
within $\sim$a few $r_\mathrm{S}$, while most of the luminosity originates
from $r \sim 5$--$50\,r_\mathrm{S}$, where the Newtonian approximation
is adequate.  The gravitational redshift correction applied to the luminosity
integral (Section~\ref{sec:efficiency}) captures the leading-order GR effect
on the observed luminosity.  We use an ideal gas equation of state with
$\gamma = 5/3$; at temperatures $T \gtrsim 10^{10}$\,K, radiation pressure
becomes non-negligible and the effective adiabatic index decreases toward
$4/3$, but this occurs only in the innermost region of the Hot Bondi
regime where bremsstrahlung cooling is already a small perturbation.
In the bremsstrahlung cooling regime ($M_\mathrm{BH} \sim 10^{-13.5}\,M_\odot$),
peak temperatures of $\sim$$10^9$\,K yield $\gamma_\mathrm{eff} \approx 1.45$--$1.55$,
a $\sim$10\% departure from $5/3$ that may modestly affect the depth of the
$\eta$ minimum; at higher masses where the flow becomes isothermal, $T \approx T_\infty$ and radiation pressure is negligible ($P_\mathrm{rad}/P_\mathrm{gas} \sim 10^{-3}$). Compton cooling off the self-generated bremsstrahlung photon field is subdominant: the photon energy density near $r_\mathrm{S}$ is $\sim$$L/(4\pi r^2 c)$, an order of magnitude below the gas energy density $\sim n k_\mathrm{B} T$.
The simulation is evolved for $\sim$$10^3$--$10^4$ dynamical times until a steady state is reached, as measured by the convergence of the mass flux $\dot{M} = 4\pi r^2 \rho v$ to within $\lesssim 1\%$ across the grid. We verify that the converged profiles satisfy the steady-state equations by evaluating the mass, momentum, and energy conservation laws in integral form on the final solution. The mass flux $\dot{M}(r) = 4\pi r^2 \rho v$ is constant to within $0.1\%$ across the grid for all masses, and the cell-integrated momentum equation closes to $1$--$3\%$, confirming that the time-dependent solver relaxes to a genuine steady state.

\subsection{Opacities, Optical Depth, and Photon Trapping}
\label{sec:eddington}

\begin{figure}
  \centering
  \includegraphics[width=\columnwidth]{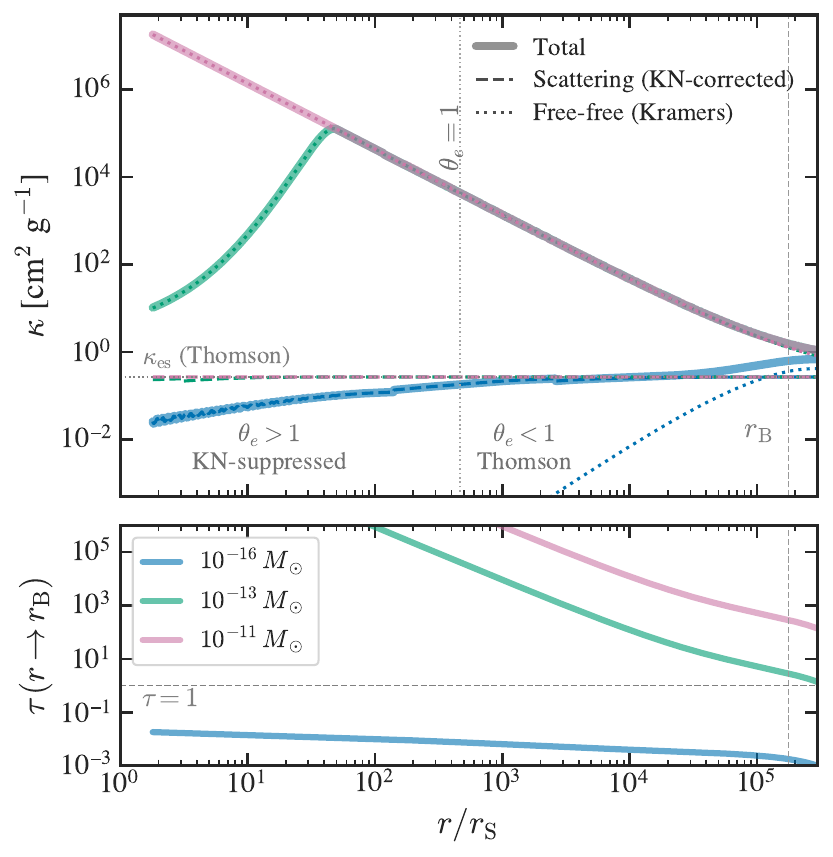}
  \caption{Opacity (top) and cumulative optical depth (bottom) as a
    function of $r/r_{\rm S}$ for $M_{\rm BH} = 10^{-16}$, $10^{-13}$,
    and $10^{-11}\,M_\odot$.  Thick translucent lines show the total
    opacity; dashed lines show Klein--Nishina--corrected scattering;
    dotted lines show Kramers free--free absorption.  The vertical
    dotted line marks the $\theta_e = 1$ boundary separating the
    KN-suppressed and Thomson regimes.  The horizontal dashed line in the
    lower panel marks $\tau = 1$.}
  \label{fig:opacity}
\end{figure}

To determine the fate of the radiation produced by the infalling gas, we
must evaluate both the local opacity and the diffusion-vs-advection
competition.  The dominant photon interactions are Thomson scattering
($\sigma_\mathrm{T}$) and free--free absorption
($\kappa_\mathrm{ff} \propto \rho T^{-7/2}$). For the Bondi profiles
($\rho \propto r^{-3/2}$, $T \propto r^{-1}$),
$\kappa_\mathrm{ff} \propto r^{2}$: free--free absorption decreases inward
because the steep temperature rise overwhelms the density increase. At
high photon energies ($h\nu \gtrsim m_e c^2$), the scattering
cross-section enters the Klein--Nishina regime,
$\sigma_\mathrm{KN} \sim \sigma_\mathrm{T}\,(\ln\theta_e)/\theta_e$,
suppressing radiation--matter coupling by a factor of $\sim$80 at
$\theta_e \sim 200$. Figure~\ref{fig:opacity} shows the radial opacity
structure for three representative BH masses.

The Bondi sphere becomes optically thick once the cumulative optical depth
$\tau(r_\mathrm{B}) \gtrsim 1$. The photon mean free path in the solar
core is $\ell_\mathrm{mfp} = 1/(n_e \sigma_\mathrm{T}) \approx 0.025$\,cm,
so $\tau \equiv r_\mathrm{B}/\ell_\mathrm{mfp} \gtrsim 1$ defines a
critical mass
\begin{equation}
  M_\mathrm{crit}^{(\tau)} = \frac{c_\infty^2\,\ell_\mathrm{mfp}}{G}
  \approx 5 \times 10^{-13}\,M_\odot.
  \label{eq:Mcrit_tau}
\end{equation}
Including the density enhancement within $r_\mathrm{B}$ and
Klein--Nishina suppression, the cumulative optical depth from
$r_\mathrm{S}$ to $r_\mathrm{B}$ reaches unity near
$M_\mathrm{BH} \sim 10^{-14}\,M_\odot$ (Figure~\ref{fig:opacity},
lower panel).  For $M_\mathrm{BH} \lesssim 10^{-14}\,M_\odot$ the
Bondi sphere is optically thin, and photons produced near $r_\mathrm{S}$
escape the accretion region without significant interaction.

For optically thick flows ($\tau(r_\mathrm{B}) \gtrsim 1$) we adopt the
spherical photon-trapping framework of \citet{Begelman1979}.  In a
strictly spherical flow, the trapping radius
\begin{equation}
  r_\mathrm{trap} \equiv \frac{\dot{M}\,\kappa}{4\pi c}
  \label{eq:rtrap}
\end{equation}
sets the boundary between an outer, diffusive zone (photons escape) and
an inner, advective zone (photons are swept inward with the flow); $\kappa$
is the relevant scattering opacity.  As $\dot{M}$ grows, $r_\mathrm{trap}$
moves outward from $r_\mathrm{S}$.  Locally generated luminosity originating
at radii $r<r_\mathrm{trap}$ is advected back to the black hole rather than
unbinding the inflow, while photons emitted further out can still diffuse
to infinity.  The escaping luminosity is then bounded by
\begin{equation}
  L_\mathrm{esc} \lesssim 0.6\,L_\mathrm{Edd}\,,
  \label{eq:Lesc_cap}
\end{equation}
where $L_\mathrm{Edd} = 4\pi G M_\mathrm{BH} m_\mathrm{p} c / \sigma$ and
$\sigma$ is the appropriate (Thomson or Klein--Nishina) cross-section;
the $\sim$0.6 prefactor is the model-dependent self-regulation value
quoted by \citet{Begelman1979}.  Critically, $\dot{M}$ continues to scale
as $\dot{M} \propto M_\mathrm{BH}^2$ in this regime --- there is no
spherical Eddington cap on the accretion rate.  The radiation pressure
exerted by the escaping component reduces the effective gravity seen by
the outer flow by a factor $(1-L_\mathrm{esc}/L_\mathrm{Edd})$, which we
incorporate self-consistently in Section~\ref{sec:efficiency}.

When $r_\mathrm{trap} \geq r_\mathrm{B}$ (at
$M_\mathrm{BH} \sim 10^{-9}\,M_\odot$, where
$r_\mathrm{trap} = \dot{M}_\mathrm{B}\,\kappa/(4\pi c) = r_\mathrm{B}$
defines the boundary), no photons escape the Bondi sphere radiatively.
Energy can still be transported outward by convection just outside
$r_\mathrm{B}$, at most at the saturated convective rate
$L_\mathrm{conv,sat} = 4\pi r_\mathrm{B}^{2}\,\rho_\infty c_\infty^{3}$
\citep{Flammang:1984}.  This is the limit usually called
``hyper-critical'' accretion in the literature
\citep{Park:1990}; in our scheme it is the full-trapping limit of the
photon-trapping regime, not a separate stage.

Spherical accretion onto a non-rotating compact object therefore admits no
Eddington-limited phase: trapping switches on continuously starting at
$\tau(r_\mathrm{B}) \sim 1$ and is complete by
$r_\mathrm{trap} = r_\mathrm{B}$.  The story differs only when an
accretion disk forms ($r_\mathrm{circ}>r_\mathrm{ISCO}$): viscous torques
then transport energy radially outward and the disk surface can radiate
at $\sim L_\mathrm{Edd}$, limiting $\dot{M}$ \citep[as in standard thin-disk
theory; see][]{Gottlieb2026}.  For the spherical solar-core problem
considered here, $r_\mathrm{circ}\ll r_\mathrm{ISCO}$ throughout the
mass range of interest.

\subsection{Radiative Feedback on the Bondi Boundary Conditions}
\label{sec:radiative_feedback}

\begin{figure}
  \centering
  \includegraphics[width=\columnwidth]{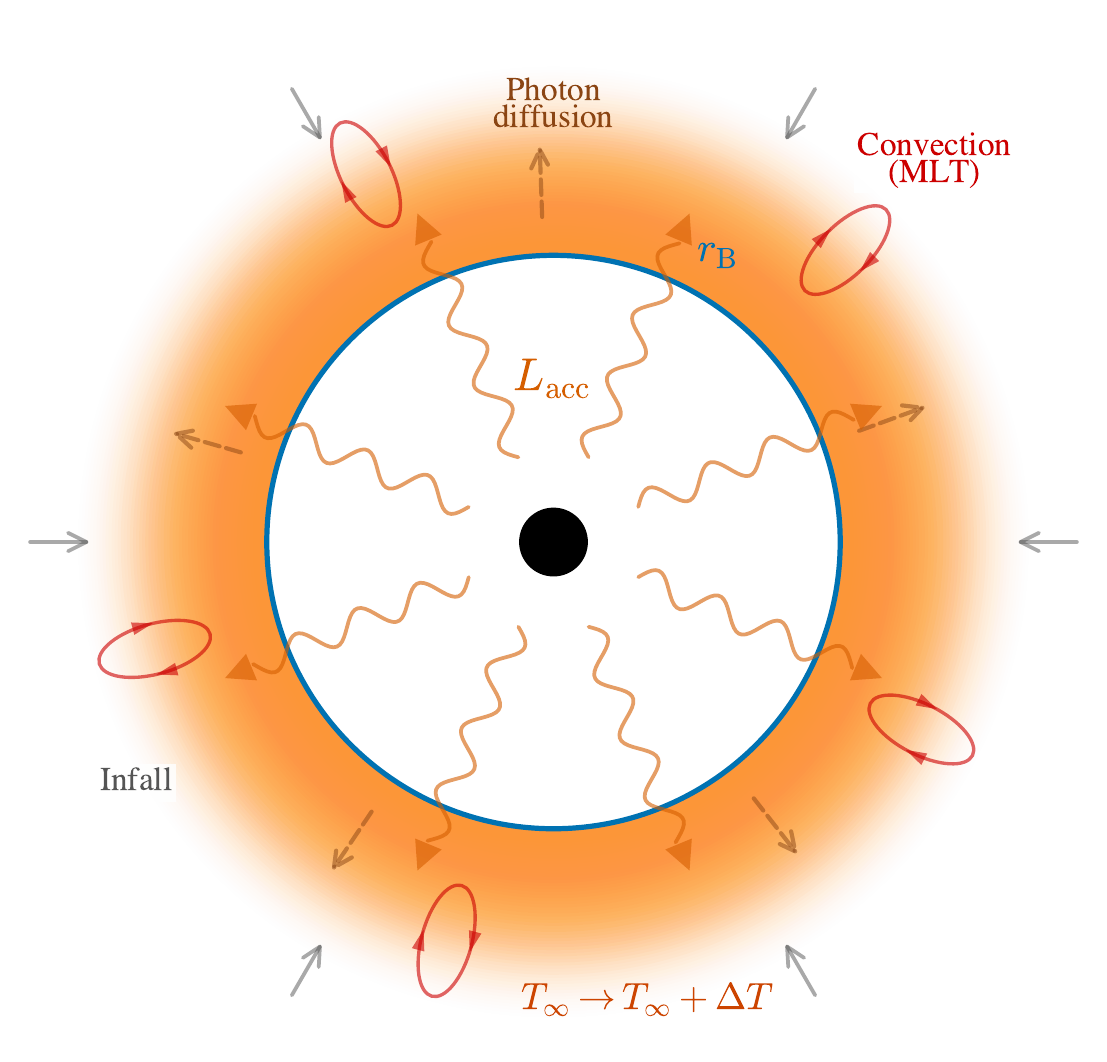}
  \caption{Schematic of the radiative feedback mechanism on the Bondi
    boundary conditions.  Accretion luminosity $L_{\rm acc}$ produced near
    the PBH (central black circle) escapes the Bondi sphere
    (blue circle, radius $r_{\rm B}$) and is absorbed by the ambient
    stellar plasma just outside it, raising the local temperature from
    $T_\infty$ to $T_\infty + \Delta T$ (orange shading).  The deposited
    energy is transported outward by photon diffusion (brown dashed arrows)
    and when the radiative gradient exceeds the adiabatic gradient, by
    convection modeled via mixing-length theory (red circulation loops).
    Grey arrows indicate the infalling gas that feeds the Bondi flow.
    The resulting temperature enhancement reduces the effective sound speed
    contrast and moderates the accretion rate.}
  \label{fig:schematic_feedback}
\end{figure}

A separate self-consistency consideration arises from the fate of the luminosity outside the Bondi sphere. In the optically thin regime, photons escape $r_\mathrm{B}$ and are absorbed by the ambient stellar plasma at a coupling radius $r_c \sim \ell_\mathrm{mfp}$.  For the hard bremsstrahlung photons that dominate at low masses ($h\nu \sim 0.1$--$1$\,MeV, Klein--Nishina regime), the mean free path
$\ell_\mathrm{mfp} \approx 1/(\kappa_\mathrm{KN}\,\rho_\infty) \approx
0.06$\,cm is a factor of $\sim$1000 larger than $r_\mathrm{B}$ for
$M_\mathrm{BH} = 10^{-15}\,M_\odot$.  The BH luminosity dominates
the local energy budget at $r_c$, since the nuclear luminosity within
this radius is negligible ($L_\mathrm{nuc} \sim 10^{-3}$\,erg\,s$^{-1}$).
Figure~\ref{fig:schematic_feedback} illustrates the feedback geometry.

The deposited energy raises the gas temperature around the BH, modifying
the effective $T_\infty$ seen by the Bondi flow.  Treating the BH as a
point source in a uniform medium, the radiative diffusion equation gives
\begin{equation}
  T^4(r) = T_\infty^4 + \frac{3\kappa\rho_\infty\,L_\mathrm{BH}}
  {4\pi a c\,r}\,,
  \label{eq:T_feedback}
\end{equation}
where $a$ is the radiation constant. Evaluating at 
$r_c = 1/(\kappa\rho_\infty)$ and defining
$x \equiv \widetilde{T}_\infty / T_\infty$, the above equation becomes
\begin{equation}
  x^4 = 1 + \beta\,x^{-3/2}\,,
  \label{eq:feedback_selfconsistent}
\end{equation}
where we have used
$L_\mathrm{BH} \propto \dot{M}_\mathrm{B} \propto c_s^{-3} \propto
\widetilde{T}_\infty^{\,-3/2}$, and the single dimensionless parameter
\begin{equation}
  \beta \equiv \frac{3(\kappa\rho_\infty)^2\,L_0}
  {4\pi a c\,T_\infty^4}
  \label{eq:beta_feedback}
\end{equation}
measures the importance of feedback, with
$L_0 = \eta\,\dot{M}_\mathrm{B}(T_\infty)\,c^2$ the unperturbed
BH luminosity.  For $\beta \ll 1$, $x \approx 1$ and feedback is negligible;
for $\beta \gg 1$, $x \approx \beta^{2/11}$ and the temperature is
significantly elevated.

In the Hot Bondi regime, the Klein--Nishina opacity
$\kappa_\mathrm{KN} \approx 0.11$\,cm$^2$\,g$^{-1}$ gives
$\beta \lesssim 0.17$ for
$M_\mathrm{BH} \leq 10^{-15}\,M_\odot$, corresponding to negligible
temperature enhancement ($\lesssim$5\%).  At higher masses
($M_\mathrm{BH} \gtrsim 10^{-14}\,M_\odot$), the emitted photons are
softer ($h\nu \sim 1$\,keV) and the Rosseland mean opacity
$\kappa \sim 1$\,cm$^2$\,g$^{-1}$ gives $\beta \gg 1$.  However, the
feedback is strongly self-regulating: an increase in $\widetilde{T}_\infty$
reduces $\eta$ and the accretion rate, lowering $L_\mathrm{BH}$.
Moreover, when $\beta \gg 1$ the radiative temperature gradient exceeds
the adiabatic gradient, triggering convective instability.  Convection
replaces the steep radiative profile $T \propto r^{-1/2}$ with a nearly
adiabatic stratification, dramatically limiting the temperature enhancement.

We quantify this by integrating a 1D hydrostatic envelope under BH
gravity using standard mixing length theory (MLT) with $\alpha_\mathrm{MLT}
= 1.5$, solving for the temperature gradient $\nabla$
such that $F_\mathrm{rad}(\nabla) + F_\mathrm{conv}(\nabla) = L/(4\pi r^2)$.
The envelope is integrated inward from $r = 200\,r_\mathrm{B}$ (where
$T = T_\infty$) to $\max(r_\mathrm{B},\,r_c)$.  The MLT model gives
temperature enhancements $x \approx 1.6$--$5$ across the cooling regime,
far smaller than the pure-diffusion prediction ($x \approx 3$--$18$),
because convection carries 30--97\% of the outward flux depending on mass.
The temperature enhancement peaks near $10^{-12}\,M_\odot$ and actually decreases at higher masses as convection becomes more efficient.

The net effect of radiative feedback is modest: accretion rates are reduced by factors of a few relative to the unperturbed Bondi rate, rather than by orders of magnitude as the pure-diffusion $\beta$ parameter might suggest. This convective saturation connects smoothly to the super-Eddington regime, where the envelope is already convective and photon trapping determines the accretion rate. The results are summarized in Figure~\ref{fig:eta}
(Section~\ref{sec:efficiency}).

\section{Results}
\label{sec:results}

\subsection{Accretion Regimes}
\label{sec:regimes}

\begin{figure*}
\centering
\includegraphics[width=\textwidth]{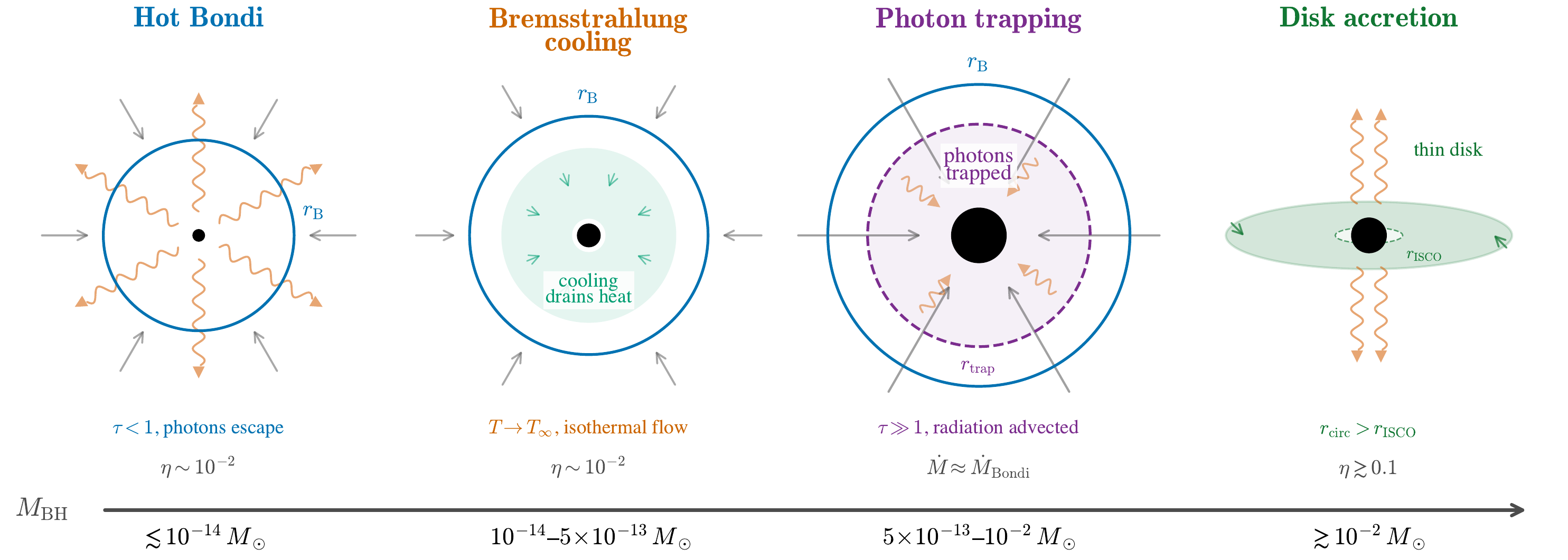}
\caption{Schematic overview of the three spherical-accretion regimes
  and the disk-accretion regime onto a PBH embedded in a stellar
  interior, ordered by increasing $M_\mathrm{BH}$.
  \textit{Hot Bondi} ($M_\mathrm{BH} \lesssim 10^{-14}\,M_\odot$):
  the Bondi sphere is optically thin ($\tau < 1$); bremsstrahlung photons
  (orange wavy arrows) escape freely without interacting with the
  infalling gas (grey arrows), yielding $\eta \approx 10^{-2}$.
  \textit{Bremsstrahlung cooling}
  ($10^{-14}$--$5\times 10^{-13}\,M_\odot$): bremsstrahlung cooling
  (green region) drains thermal energy before the gas reaches
  pair-producing temperatures; the flow approaches isothermal
  ($T \to T_\infty$) and $\eta \approx 10^{-2}$.
  \textit{Photon trapping}
  ($5\times 10^{-13}\lesssim M_\mathrm{BH}/M_\odot \lesssim 10^{-2}$):
  the Bondi sphere is optically thick; in spherical geometry the trapping
  radius $r_\mathrm{trap}$ moves outward from $r_\mathrm{S}$ and most
  locally produced photons are advected inward rather than escaping
  \citep{Begelman1979}.  The escaping luminosity is bounded by
  $\sim 0.6\,L_\mathrm{Edd}$ but does not cap the accretion rate, and
  $\dot{M} \approx \dot{M}_\mathrm{B}$ throughout.
  \textit{Disk accretion}
  ($M_\mathrm{BH} \gtrsim 10^{-2}\,M_\odot$, depending on stellar
  rotation): once $r_\mathrm{circ}>r_\mathrm{ISCO}$ a thin disk forms,
  $\eta \approx 0.08$, and the standard Eddington limit applies via
  viscous angular-momentum transport \citep{Gottlieb2026}.}
\label{fig:schematic}
\end{figure*}

As $M_\mathrm{BH}$ increases, the accretion flow passes through three
physically distinct regimes for spherical accretion (Hot Bondi,
Bremsstrahlung cooling, Photon trapping), each with a characteristic
radiative efficiency $\eta \equiv L / \dot{M} c^2$. A fourth regime,
Disk accretion, opens once $r_\mathrm{circ}>r_\mathrm{ISCO}$ and the
spherical treatment no longer applies; this transition lies above the
mass range we resolve dynamically and is discussed in our companion
paper \citep{Gottlieb2026}. Figure~\ref{fig:schematic} provides a
schematic overview.

\begin{figure*}
\centering
\includegraphics[width=\textwidth]{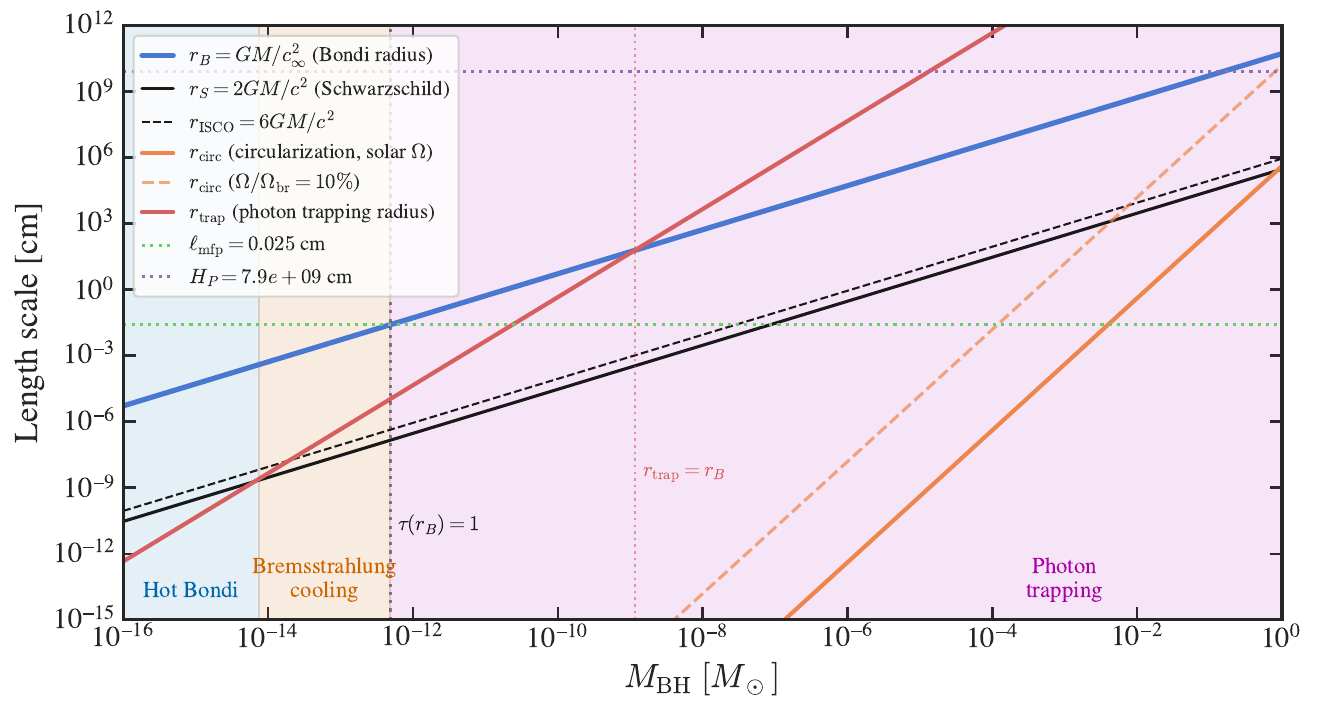}
\caption{Physical length scales controlling the accretion regime as a
  function of PBH mass, for solar core conditions.  The Bondi radius
  $r_\mathrm{B}$ (blue) defines the gravitational sphere of influence.
  The vertical dotted line marks $\tau(r_\mathrm{B}) = 1$, where the
  Bondi sphere becomes optically thick
  ($M \sim 5 \times 10^{-13}\,M_\odot$).  The photon trapping
  radius $r_\mathrm{trap}$ (red) exceeds $r_\mathrm{B}$ at
  $M \sim 10^{-9}\,M_\odot$.  Two circularization radii
  $r_\mathrm{circ}$ (orange) are shown: one for the present-day solar
  core rotation ($\Omega/\Omega_\mathrm{br} \approx 5 \times 10^{-4}$,
  solid) and one for 10\% of the core break-up frequency
  $\Omega_\mathrm{br} = \sqrt{GM_\mathrm{core}/r_\mathrm{core}^3}
  \approx 5.3 \times 10^{-3}$\,rad\,s$^{-1}$ (dashed).
  An accretion disk forms where $r_\mathrm{circ}$ crosses
  $r_\mathrm{ISCO}$ (black dashed).  Young solar-type stars
  rotate significantly faster than the present Sun, so disk formation
  may occur at lower BH masses during the early evolution.
  Colored bands denote the three spherical-accretion regimes discussed in
  Section~\ref{sec:regimes}; the onset of trapping at $\tau(r_\mathrm{B})=1$
  marks the boundary into the photon-trapping regime.}
\label{fig:scales}
\end{figure*}

Figure~\ref{fig:scales} shows the relevant physical length scales as a
function of $M_\mathrm{BH}$, with the regime boundaries indicated.

\begin{figure*}
  \centering
  \includegraphics[width=\textwidth]{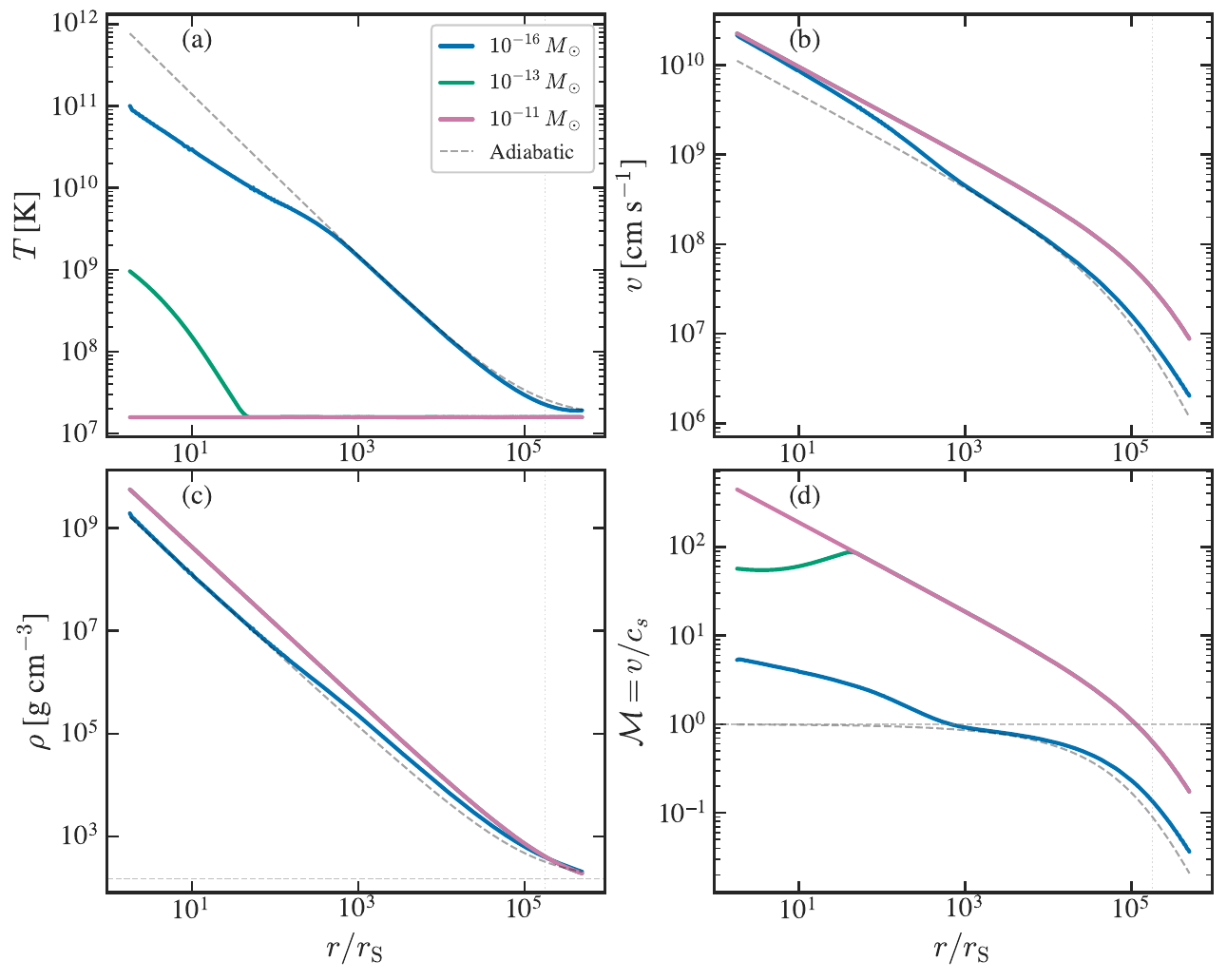}
  \caption{Steady-state radial profiles from the time-dependent Euler
    solver for $M_{\rm BH} = 10^{-16}$, $10^{-13}$, and
    $10^{-11}\,M_\odot$.
    (a)~Temperature, (b)~infall velocity, (c)~density, and
    (d)~Mach number as a function of $r/r_{\rm S}$.  Dashed grey lines
    show the adiabatic Bondi solution.  The vertical dashed line marks
    $r_{\rm B}$.  The transition from near-adiabatic (low mass) to
    nearly isothermal (high mass) flow is evident in the temperature
    panel, while the density and velocity profiles show the enhanced
    accretion rate in the isothermal limit.}
  \label{fig:profiles}
\end{figure*}

\begin{enumerate}
\item \textbf{Hot Bondi regime} ($M_\mathrm{BH} \lesssim 10^{-14}\,M_\odot$; $\eta \approx 0.01$--$0.1$, peaking near $10^{-15}\,M_\odot$).  Bremsstrahlung cooling is dynamically negligible ($t_\mathrm{cool}/t_\mathrm{ff}\gg 1$), so the density and velocity follow the adiabatic Bondi solution and the gas heats to $T\sim 10^{11}$\,K near $r_\mathrm{S}$.  At those temperatures, the Coulomb mean free path exceeds the local flow scale (Appendix~\ref{sec:collisionless_pileup}) before the gas reaches $\theta_e \sim 1$. Beyond this point the infalling particles are essentially free-streaming: they gain kinetic energy from the gravitational potential but cannot thermalize it through collisions. Without a thermalized particle distribution there is no thermal photon bath, and the equilibrium pair production channel ($\gamma\gamma \to e^+e^-$) is suppressed. Collisional (``trident'') pair production ($e + Z \to e + Z + e^+e^-$) remains possible in principle, but its cross-section is smaller than the bremsstrahlung cross-section by factors of $\sim$$10^6$--$10^9$ and is negligible. Bremsstrahlung, which requires only individual binary encounters rather than full thermalization, dominates the cooling. The adiabatic Bondi solution is an excellent approximation to the velocity and density profiles, and the luminosity is produced by relativistic bremsstrahlung from a region near $r_\mathrm{S}$. A subtlety of this regime is that individual particles inside the collisionless zone retain their thermal angular momentum, so most have $\ell \gg \ell_\mathrm{crit} = 4 G M_\mathrm{BH}/c$ and miss the general-relativistic loss cone on a single pass. Recycling through the collisional outer flow, combined with bremsstrahlung-driven circularization and self-induced collisionalization of the resulting density pile-up, limits the correction to $\dot{M}$ to at most $\sim 30\%$ at $10^{-16}\,M_\odot$ and to a few percent or less for $M_\mathrm{BH} \gtrsim 10^{-15.8}\,M_\odot$; plasma microinstabilities in the counter-streaming inner flow likely add anomalous collisionality and push the correction further toward unity. We give the full argument in Appendix~\ref{sec:collisionless_pileup}.

\item \textbf{Bremsstrahlung cooling regime} ($10^{-14} \lesssim M_\mathrm{BH}/M_\odot \lesssim 5\times10^{-13}$; $\eta \approx 10^{-2}$). At fixed dimensionless radius $x = r/r_\mathrm{B}$, the adiabatic Bondi profiles of $\rho$, $v$, and $T$ (and hence the local emissivity) are independent of $M_\mathrm{BH}$ (since $r_\mathrm{B}/r_\mathrm{S} = c^2/2c_\infty^2$ is mass-independent). What changes is the physical size of the Bondi sphere: $r_\mathrm{B} \propto M_\mathrm{BH}$, so the infall time $t_\mathrm{ff} \propto M_\mathrm{BH}$ at fixed $x$, while the cooling time $t_\mathrm{cool} \sim n k_\mathrm{B} T / \varepsilon$ is mass-independent. The ratio $t_\mathrm{cool}/t_\mathrm{ff} \propto M_\mathrm{BH}^{-1}$ therefore decreases with mass: cooling is a negligible perturbation at $10^{-16}\,M_\odot$ ($t_\mathrm{cool}/t_\mathrm{ff} \sim 10^2$), but by $10^{-14}\,M_\odot$ it drains thermal energy before the gas reaches $\theta_e \sim 1$. Pairs are therefore preempted by cooling, not suppressed by collisionlessness as in the previous regime. % The temperature profile flattens toward $T \approx T_\infty$, while the density and velocity profiles remain nearly adiabatic. The radiative efficiency $\eta$ initially \emph{decreases} because $T_\mathrm{max}$ drops from $\sim 10^{10}$\,K to $\sim T_\infty$, reducing the bremsstrahlung emissivity ($\varepsilon_\mathrm{ff} \propto T^{1/2}$ at fixed $\rho$), while the reduced thermal pressure support simultaneously enhances $\dot{M}$. Both effects lower $\eta$, which reaches a minimum $\eta \approx 10^{-2}$ near $M_\mathrm{BH} \sim 3 \times 10^{-13}\,M_\odot$. At higher masses the flow becomes nearly isothermal ($T \approx T_\infty$) and $\dot{M}$ saturates at $\sim$7 times the adiabatic Bondi rate; $\eta$ then rises with mass as the growing Bondi sphere produces more bremsstrahlung luminosity ($L \propto r_\mathrm{B}^3 \propto M^3$, while $\dot{M} \propto M^2$).

\item \textbf{Photon-trapping regime} ($5\times 10^{-13} \lesssim M_\mathrm{BH}/M_\odot \lesssim 10^{-2}$).
The Bondi sphere becomes optically thick at $M_\mathrm{BH} \sim 5\times 10^{-13}\,M_\odot$ (Section~\ref{sec:eddington}).
In spherical geometry the trapping radius $r_\mathrm{trap} = \dot{M}\kappa/(4\pi c)$ moves outward continuously
from $r_\mathrm{S}$ as $\dot{M}$ grows; locally produced photons interior to $r_\mathrm{trap}$ are advected back
to the black hole, while only the outer diffusive zone radiates to infinity \citep{Begelman1979}.  The escaping
luminosity self-regulates at $L_\mathrm{esc} \lesssim 0.6\,L_\mathrm{Edd}$, but $\dot{M}$ continues to follow the
Bondi rate: there is no spherical Eddington cap.  The radiation pressure exerted by the escaping component
reduces the effective gravity seen by the outer flow by a factor $(1-L_\mathrm{esc}/L_\mathrm{Edd})$, smoothly
modulating $\dot{M}$ between the isothermal-limit enhancement ($\sim 7\,\dot{M}_\mathrm{B}$) and the adiabatic
Bondi rate.  Full trapping ($r_\mathrm{trap}\geq r_\mathrm{B}$) is reached near $M_\mathrm{BH} \sim 10^{-9}\,M_\odot$,
beyond which the escaping component itself is suppressed and $\dot{M}\to\dot{M}_\mathrm{B}$ exactly; convective
transport at the Bondi boundary continues to carry a residual luminosity $L_\mathrm{conv,sat} = 4\pi r_\mathrm{B}^2
\rho_\infty c_\infty^3$ \citep{Flammang:1984}.  Throughout the regime, $\dot{M}\propto M^2$ and growth remains
super-exponential, with $t_\mathrm{B}\propto M_\mathrm{BH}^{-1}$ giving
$M(t) = M_0/(1 - t/t_\mathrm{B})$.

\end{enumerate}
In our companion paper, \citet{Gottlieb2026}, we show that at sufficiently high $M_\mathrm{BH}$, the circularization radius $r_\mathrm{circ} = j^2 / (G M_\mathrm{BH})$ (where $j = \Omega r^2$ is the specific angular momentum of the stellar core) exceeds the innermost stable circular orbit $r_\mathrm{ISCO} = 6\,G M_\mathrm{BH}/c^2$, and an accretion disk forms. For the present-day solar core rotation ($\Omega / \Omega_\mathrm{br} \approx 5 \times 10^{-4}$), this requires $M_\mathrm{BH} \sim 1.5\,M_\odot$. Young solar-type stars rotate much faster: at 1\% of break-up (typical of a $\sim$1\,Gyr solar analog), disk formation occurs at $\sim$$0.08\,M_\odot$, and at 10\% (early pre-main-sequence) at $\sim$$0.008\,M_\odot$ (Figure~\ref{fig:scales}). Below the disk-formation mass, angular momentum is dynamically unimportant and spherical accretion is appropriate. Figure~\ref{fig:profiles} shows the steady-state radial profiles of temperature, velocity, density, and Mach number from the time-dependent Euler solver for three representative masses spanning the Hot Bondi and bremsstrahlung cooling regimes. For the lowest mass ($10^{-16}\,M_\odot$), all profiles closely follow the adiabatic Bondi solution: bremsstrahlung cooling is a small perturbation, and the flow reaches $T \sim 10^{11}$\,K near $r_\mathrm{S}$. At $10^{-13}\,M_\odot$, cooling flattens the temperature profile and the gas never heats above $\sim 10^{9}$\,K, while the density and velocity remain nearly adiabatic. At $10^{-11}\,M_\odot$, the flow is nearly isothermal ($T \approx T_\infty$) throughout, with the accretion rate enhanced to $\sim$7$\,\dot{M}_\mathrm{B}$ and Mach numbers reaching $\mathcal{M} \sim 100$ at the inner boundary.

%Note that the flow properties do not match the ambient stellar-interior values ($T_\infty$, $\rho_\infty$) at $r_\mathrm{B}$: the Bondi radius marks where the gravitational and thermal energies are comparable, but the flow is already accreting there, with $T(r_\mathrm{B}) \approx \tfrac{5}{4}\,T_\infty$ and $\rho(r_\mathrm{B}) \approx 1.4\,\rho_\infty$ for $\gamma = 5/3$. The unperturbed ambient conditions are recovered only at $r \gg r_\mathrm{B}$. For the isothermal case ($10^{-11}\,M_\odot$), the enhanced accretion rate ($\dot{M} \approx 7\,\dot{M}_\mathrm{B}$) amplifies this offset, as the velocity and density profiles are systematically above the adiabatic Bondi values throughout the computational domain.

\subsection{Radiative Efficiency and Luminosity} \label{sec:efficiency} We compute the radiative efficiency $\eta(M_\mathrm{BH})$ from the time-dependent simulations by evaluating the total luminosity 
$L = \int_{r_\mathrm{S}}^{r_\mathrm{B}} 4\pi r^2 \varepsilon(\rho, T)\,dr$ 
on the converged steady-state profiles and dividing by $\dot{M} c^2$. A
gravitational redshift correction $(1 - r_\mathrm{S}/r)$ is applied to
the integrand; the net effect on $\eta$ is $\sim$5\%, since most of the
luminosity originates from $r \sim 5$--$50\,r_\mathrm{S}$. To incorporate
radiative feedback, we construct a composite $\eta(M_\mathrm{BH})$ from
the diffusion feedback model (Section~\ref{sec:radiative_feedback}) where
$\beta < 1$ and the MLT convective envelope model where $\beta \geq 1$.
Analytical fitting formulae for $\eta(M)$ and $f(M)$ are given in
Appendix~\ref{sec:analytical_fits}.

\begin{figure*}
\centering
\includegraphics[width=\textwidth]{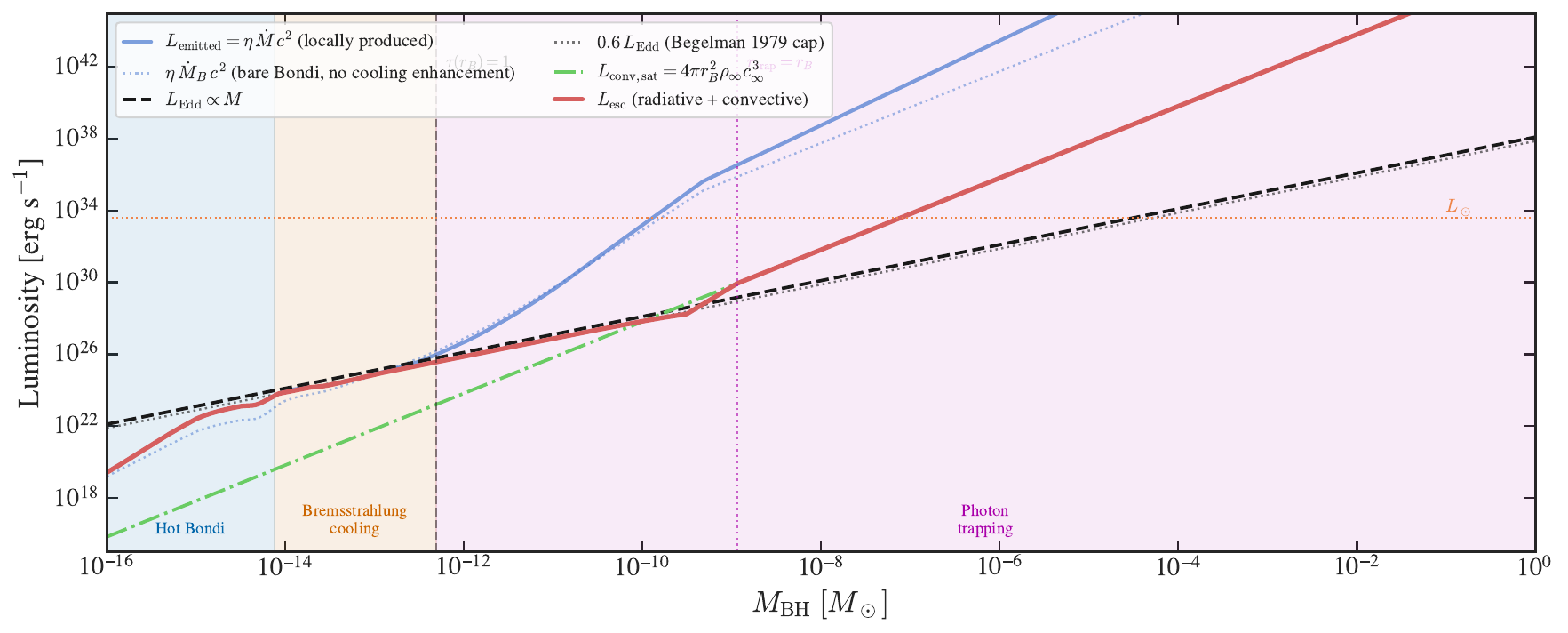}
\caption{Accretion luminosity as a function of BH mass.  The solid blue
  line shows the locally emitted luminosity $L_\mathrm{emitted} = \eta\,\dot{M}\,c^2$
  evaluated with the cooling-enhanced accretion rate; the dotted blue line shows
  the bare-Bondi reference $\eta\,\dot{M}_\mathrm{B} c^2$ without the cooling
  enhancement.  The black dashed line is $L_\mathrm{Edd} \propto M$, the dotted
  black line is the \citet{Begelman1979} cap $0.6\,L_\mathrm{Edd}$, and the
  green dash-dotted line is the saturated-convection floor $L_\mathrm{conv,sat}
  = 4\pi r_\mathrm{B}^2 \rho_\infty c_\infty^3$.  The red thick line is the
  physically escaping luminosity $L_\mathrm{esc}=\max\bigl(f_\mathrm{geom}
  \min(L_\mathrm{emitted}, 0.6\,L_\mathrm{Edd}),\,
  (1-f_\mathrm{geom})\,L_\mathrm{conv,sat}\bigr)$, with
  $f_\mathrm{geom} = 1 - r_\mathrm{trap}/r_\mathrm{B}$; it is capped by the
  Begelman bound in the partial-trapping window and asymptotes to
  $L_\mathrm{conv,sat}$ once $r_\mathrm{trap} \geq r_\mathrm{B}$.  Background shading marks the three accretion regimes.  The
  solar luminosity $L_\odot$ is shown for reference.}
\label{fig:luminosity}
\end{figure*}

Figure~\ref{fig:eta} shows $\eta(M_\mathrm{BH})$ from the simulations.
At low masses ($M_\mathrm{BH} \lesssim 5\times 10^{-13}\,M_\odot$),
$\eta \approx 10^{-2}$; in the photon-trapping regime above
$M_\mathrm{BH}\sim 5\times 10^{-13}\,M_\odot$ the locally emitted
luminosity $L_\mathrm{emitted} = \eta\,\dot{M}\,c^2 = \eta\,f(M)\,\dot{M}_\mathrm{B}\,c^2$
from the optically thin solution rises rapidly with mass ($L\propto r_\mathrm{B}^3
\propto M^3$ while $\dot{M}\propto M^2$, giving $\eta\propto M$ in the
nominal fit), but the physical escape is capped by the
\citet{Begelman1979} bound $L_\mathrm{esc}\lesssim 0.6\,L_\mathrm{Edd}$
(Section~\ref{sec:eddington}).  The plotted $\eta$ values in this regime
should be interpreted as the locally emitted efficiency before trapping,
not as the externally observed efficiency; the latter is obtained by
substituting $L_\mathrm{esc}$ in the numerator and is shown in
Figure~\ref{fig:luminosity}.  We therefore do not use the rising
optically-thin $\eta$ for $M\gtrsim 5\times 10^{-13}\,M_\odot$ in the
growth calculations of Section~\ref{sec:growth}; instead, Equation
\eqref{eq:Mdot_actual} substitutes the effective-gravity reduction
$(1-L_\mathrm{esc}/L_\mathrm{Edd})^2$ for the Eddington cap of earlier
treatments.

The self-consistently computed efficiency is $\sim$1 order of magnitude
below the canonical thin-disk value $\eta \approx 0.06$--$0.1$ commonly
assumed in the literature \citep[e.g.,][]{Bellinger2023}.  Because the
spherical accretion rate is never Eddington-capped, there is no Salpeter
phase: the BH grows super-exponentially throughout the photon-trapping
regime.  The growth bottleneck is set entirely by the slow Hot Bondi
phase, not by an Eddington-limited stage.  Figure~\ref{fig:luminosity}
shows the resulting accretion luminosity.  In the optically thin regime,
the Eddington limit is irrelevant and $L = \eta \dot{M}_\mathrm{B} c^2$.
In the photon-trapping regime, the escaping luminosity follows the
\citet{Begelman1979} cap and transitions smoothly to the saturated
convective floor $L_\mathrm{conv,sat} = 4\pi r_\mathrm{B}^2 \rho_\infty
c_\infty^3$ when $r_\mathrm{trap} \geq r_\mathrm{B}$
(Section~\ref{sec:eddington}).

\begin{figure}
\centering
\includegraphics[width=\columnwidth]{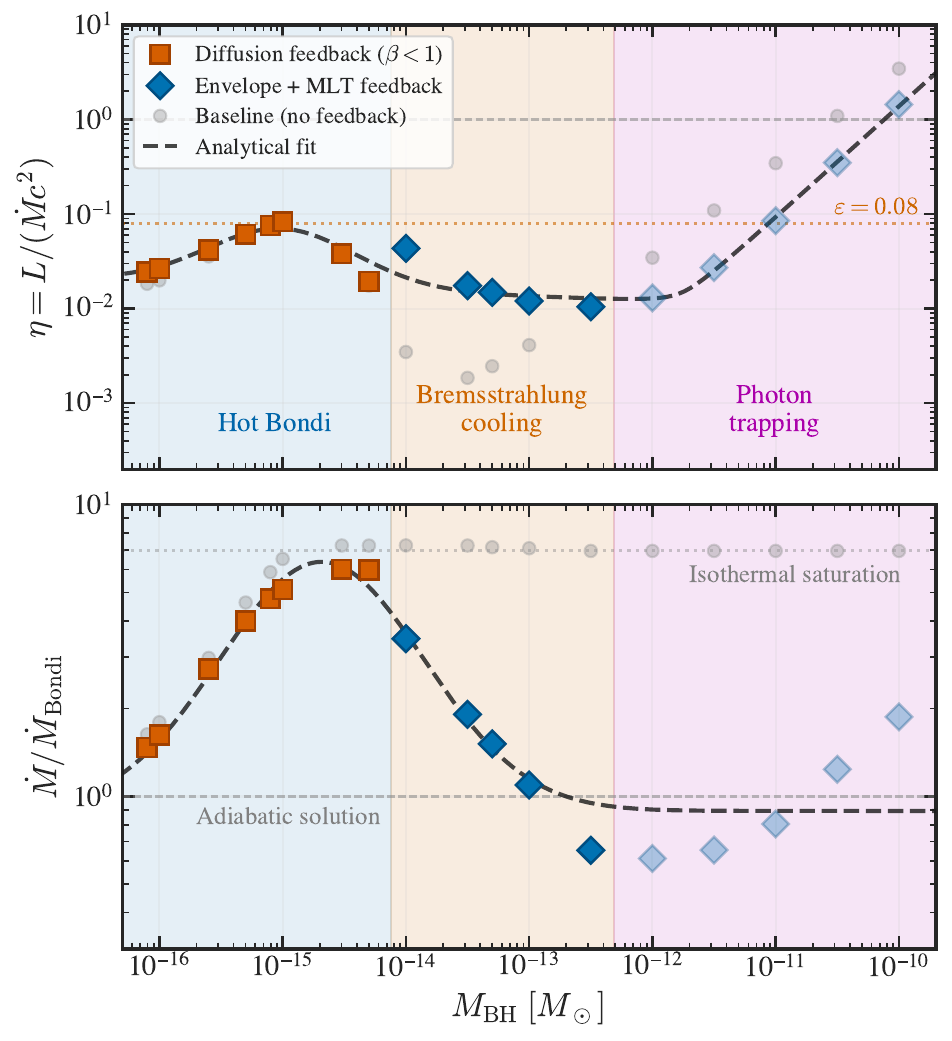}
\caption{\textit{Top:} Radiative efficiency $\eta = L/\dot{M}c^2$ as a function of PBH mass.  \textit{Bottom:} Accretion rate enhancement $\dot{M}/\dot{M}_{\rm Bondi}$ over the adiabatic Bondi rate. Vermillion squares show diffusion-feedback results (where $\beta < 1$), blue diamonds show envelope + MLT feedback results, and gray circles show the baseline (no feedback) runs.  Dashed black curves show the analytical fits (Equations~\ref{eq:eta_fit} and~\ref{eq:f_fit}).  The horizontal dotted line in the top panel marks $\eta = 0.08$ assumed by \citet{Bellinger2023}; our self-consistent efficiency is $\sim$1 order of magnitude lower across the optically thin regime. Background shading indicates the three spherical-accretion regimes: Hot Bondi (blue), bremsstrahlung cooling (orange), and photon trapping (purple). In the photon-trapping regime ($M_\mathrm{BH}\gtrsim 5\times 10^{-13}\,M_\odot$), the plotted $\eta$ values are the locally emitted efficiency from the optically thin Bondi+cooling solution; they overestimate the externally observed efficiency, which is capped by $L_\mathrm{esc}\lesssim 0.6\,L_\mathrm{Edd}$ \citep{Begelman1979}.  Equation~\eqref{eq:Mdot_actual} uses these values together with the effective-gravity reduction $(1-L_\mathrm{esc}/L_\mathrm{Edd})^2$ to compute the physical accretion rate.}
\label{fig:eta}
\end{figure}

\subsection{Growth Trajectories and Critical Initial Mass}
\label{sec:growth}

In the spherical regime the instantaneous growth rate is set by the Bondi
rate modulated by the cooling enhancement and by the radiation-pressure
reduction in effective gravity from the escaping luminosity:
\begin{equation}
  \dot{M} = \dot{M}_\mathrm{B}(M)\,
            \bigl[1 + (f(M)-1)\,\eta_\mathrm{cool}\bigr]\,
            \bigl(1 - L_\mathrm{esc}/L_\mathrm{Edd}\bigr)^2\,,
  \label{eq:Mdot_actual}
\end{equation}
where $f(M) = \dot{M}/\dot{M}_\mathrm{B}|_\mathrm{iso}$ is the cooling
enhancement factor in the isothermal limit (Figure~\ref{fig:eta},
bottom panel; Appendix~\ref{sec:analytical_fits}), ranging from $\approx$1.6
at $10^{-16}\,M_\odot$ to $\approx$7 near $10^{-14}\,M_\odot$;
$\eta_\mathrm{cool} \equiv L_\mathrm{esc}/L_\mathrm{emitted}\in[0,1]$ is
the fraction of the locally produced luminosity that actually escapes
(and therefore drives the cooling); and $L_\mathrm{esc} =
f_\mathrm{geom}(r_\mathrm{trap}/r_\mathrm{B})\,\min(L_\mathrm{emitted},
0.6\,L_\mathrm{Edd})$ is the radiative escape capped at the
\citet{Begelman1979} bound, with the geometric escape fraction
$f_\mathrm{geom}=\max(0,1-r_\mathrm{trap}/r_\mathrm{B})$ smoothly
interpolating between optically thin escape and full trapping.  In the
optically thin limit ($\tau<1$), $\eta_\mathrm{cool}\to 1$ and
$L_\mathrm{esc}\ll L_\mathrm{Edd}$, so $\dot{M}\to f(M)\,\dot{M}_\mathrm{B}$.
In the fully trapped limit ($r_\mathrm{trap}\geq r_\mathrm{B}$),
$L_\mathrm{esc}\to 0$ and $\eta_\mathrm{cool}\to 0$, so
$\dot{M}\to\dot{M}_\mathrm{B}$ (adiabatic Bondi).  In the intermediate
partial-trapping window, $L_\mathrm{esc}/L_\mathrm{Edd}$ saturates near
$0.6$ and the $(1-L_\mathrm{esc}/L_\mathrm{Edd})^2$ factor smoothly
reduces $\dot{M}$ by up to a factor $\sim 0.16$ below the adiabatic
value.  Crucially, $\dot{M}\propto M^2$ is preserved throughout: there
is no spherical Eddington plateau, and the growth remains
super-exponential with $M(t) = M_0/(1 - t/t_\mathrm{B})$,
$t_\mathrm{B}\propto M_0^{-1}$.

Once $r_\mathrm{circ}>r_\mathrm{ISCO}$ a disk forms and the rate is
instead set by the standard thin-disk Eddington limit
$\dot{M}_\mathrm{Edd}^\mathrm{disk}=L_\mathrm{Edd}/(\eta_\mathrm{disk}c^2)$
with $\eta_\mathrm{disk}\approx 0.08$; this transition lies beyond the
spherical-accretion mass range and is treated in our companion paper
\citep{Gottlieb2026}.

\begin{figure*}
\centering
\includegraphics[width=\textwidth]{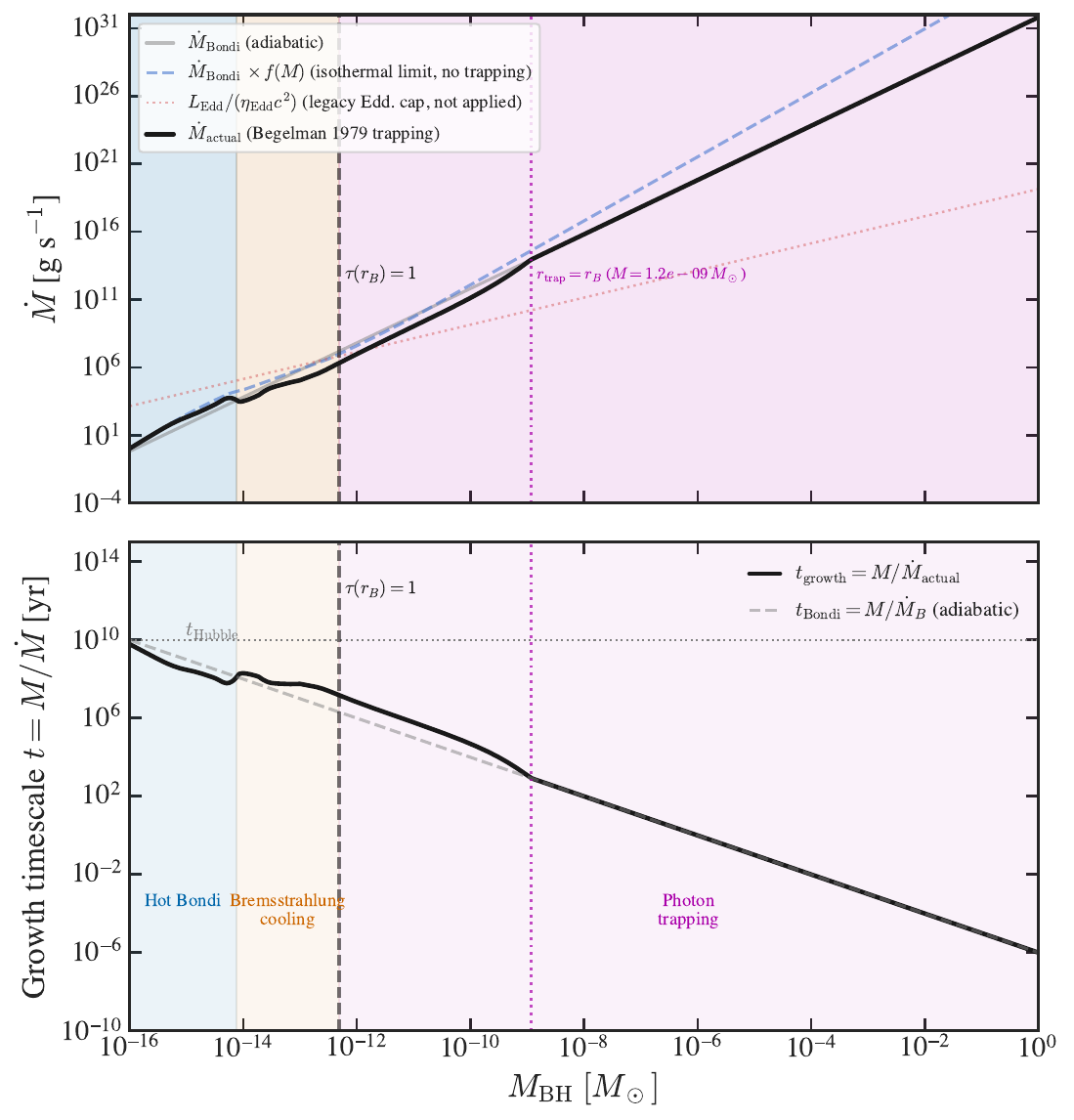}
\caption{\emph{Top panel:} Accretion rate onto a PBH in the solar core
  as a function of BH mass.  The gray solid line shows the adiabatic
  Bondi rate $\dot{M}_\mathrm{B} \propto M^2$; the blue dashed line
  shows the isothermal-limit cooling-enhanced rate $f(M)\dot{M}_\mathrm{B}$;
  the red dotted line shows the legacy Eddington cap
  $L_\mathrm{Edd}/(\eta_\mathrm{Edd} c^2)$ for reference (\emph{not}
  applied in our prescription); and the thick black line shows the
  physical accretion rate of Equation~\eqref{eq:Mdot_actual}, which
  incorporates the \citet{Begelman1979} photon-trapping framework and
  the effective-gravity reduction from the escaping luminosity.
  Vertical markers indicate the onset of trapping at
  $\tau(r_\mathrm{B})=1$ and the full-trapping mass
  ($r_\mathrm{trap} = r_\mathrm{B}$, an internal landmark, not a regime
  boundary).  Colored bands indicate the three spherical-accretion
  regimes.
  \emph{Bottom panel:} Corresponding growth timescale $t = M/\dot{M}$.
  The Hubble time is marked as a horizontal dotted line.  The growth
  time decreases monotonically with mass throughout the spherical
  regime (other than a small bump at $\mbh \sim 10^{-14}\,M_\odot$ in the transition from the Hot Bondi to Bremsstrahlung cooling regime); the partial-trapping window introduces only a smooth $O(1)$
  reduction relative to adiabatic Bondi growth, with no Salpeter
  plateau.}  
  %The Hot Bondi regime sets the bottleneck for
  %$M_\mathrm{0,crit}$.}
\label{fig:accretion}
\end{figure*}

\begin{figure*}
\centering
\includegraphics[width=\textwidth]{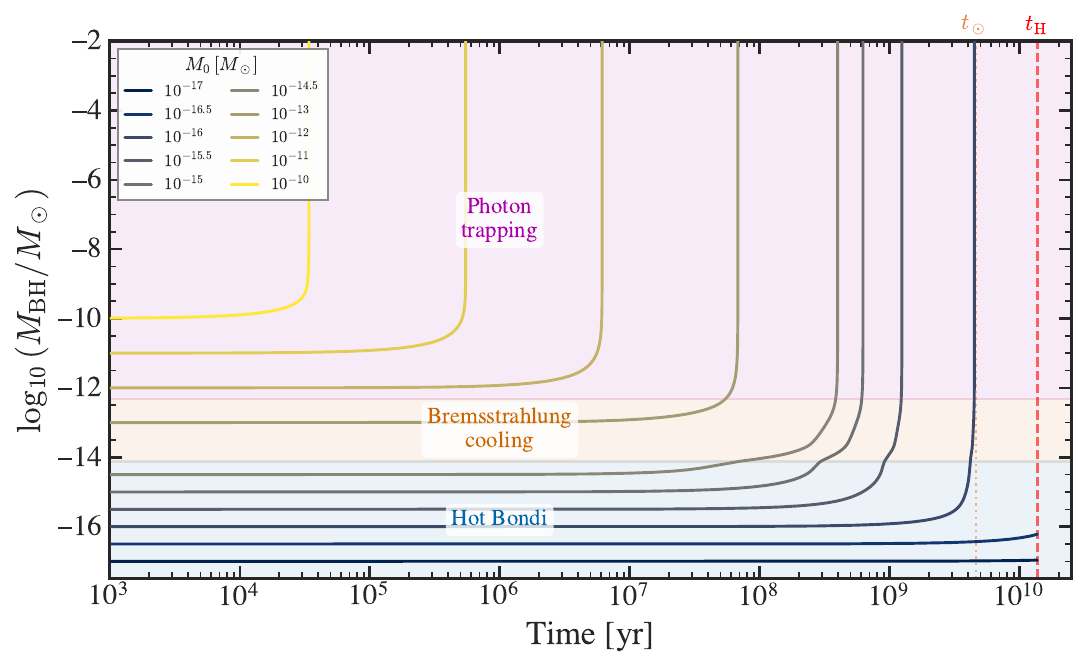}
\caption{Growth trajectories $M_\mathrm{BH}(t)$ for PBHs with initial
  masses $M_0 = 10^{-17}$--$10^{-10}\,M_\odot$ embedded in the solar
  core (color scale, from light to dark green).  Trajectories are
  trimmed at $0.01\,M_\odot$. PBHs with
  $M_0 \gtrsim 10^{-16}\,M_\odot$ reach this threshold within a Hubble
  time; lighter PBHs stall in the Hot Bondi regime.  Horizontal
  colored bands indicate the accretion regimes from
  Section~\ref{sec:regimes}.  Vertical lines mark the solar age
  ($t_\odot = 4.6$\,Gyr) and the Hubble time
  ($t_\mathrm{H} = 13.7$\,Gyr).}
\label{fig:growth}
\end{figure*}

Figure~\ref{fig:accretion} shows the resulting accretion rate and growth
timescale.  We integrate Equation~\eqref{eq:Mdot_actual} for a grid of
initial masses $M_0$ spanning $10^{-17}$--$10^{-10}\,M_\odot$. For
$M_0 \lesssim M_\mathrm{0,crit} \approx 10^{-16}\,M_\odot$, the initial
Bondi timescale exceeds a Hubble time and the PBH barely grows; for
$M_0 \gtrsim M_\mathrm{0,crit}$ the PBH reaches $0.01\,M_\odot$ within
$t_\mathrm{H}$, with the final growth stages essentially instantaneous
on cosmological timescales.  The critical mass depends on the stellar
properties (primarily $\rho_\infty$ and $c_\infty$) but is insensitive
to the target mass, since the bottleneck is set entirely by the slow
Hot Bondi phase, not by any Eddington-limited stage.
Figure~\ref{fig:growth} shows growth trajectories for representative
initial masses, illustrating how PBHs with
$M_0 \gtrsim 10^{-16}\,M_\odot$ all consume their host star within a
Hubble time.

\section{Discussion}
\label{sec:discussion}

\subsection{Magnetic Flux Accumulation and Magnetically Arrested Accretion}
\label{sec:magnetic}

A prerequisite for any magnetic effect on the accretion flow is that the
field be frozen into the plasma, which requires a magnetic Reynolds number
$\mathrm{R_m} = v\,r_\mathrm{B}/\eta_\mathrm{m} \gg 1$, where
$\eta_\mathrm{m}$ is the magnetic diffusivity.  In the radiative solar core, the relevant diffusivity is
the molecular (Spitzer) value
$\eta_\mathrm{m} \approx c^2/(4\pi\sigma) \sim 10^2$\,cm$^2$\,s$^{-1}$
for $T \approx 1.5 \times 10^7$\,K.  With $v \sim c_\infty \approx
5 \times 10^7$\,cm\,s$^{-1}$, this gives
$\mathrm{R_m} \sim 5 \times 10^5\,r_\mathrm{B}/\mathrm{cm}$, so flux freezing
holds ($\mathrm{R_m} \gg 1$) for all PBH masses of interest
($r_\mathrm{B} \gtrsim 10^{-5}$\,cm).  We note, however, that the
radiative feedback discussed in Section~\ref{sec:radiative_feedback} can
drive convection in the envelope surrounding the Bondi sphere, raising
the effective diffusivity to the turbulent value
$\eta_\mathrm{t} \sim 10^7$\,cm$^2$\,s$^{-1}$ and reducing
$\mathrm{R_m}$ to $\sim$$5\,r_\mathrm{B}/\mathrm{cm}$.  In this case,
$\mathrm{R_m} \lesssim 1$ for $M_\mathrm{BH} \lesssim$ a few
$\times\,10^{-12}\,M_\odot$, and the field resistively decouples from
the flow---making flux accumulation impossible in the first place.  For
higher masses where $\mathrm{R_m} \gg 1$ even with turbulent diffusivity,
the question reduces to whether the frozen-in field can grow to dynamical
importance.

A potential concern is that infalling plasma dragging magnetic flux onto the BH may lead to magnetically driven feedback.  If the flux threading
the Bondi sphere, $\Phi_\mathrm{B} \sim B_\star\,r_\mathrm{B}^2$, is
carried inward without expulsion, the field at radius $r$ scales as
$B(r) \sim \Phi_\mathrm{B}/r^2$ and the ratio of magnetic to ram
pressure at the gravitational radius $r_g = GM/c^2$ is
\begin{equation}
  \frac{P_\mathrm{mag}}{P_\mathrm{ram}}\bigg|_{r_g}
  \sim \frac{B_\star^2}{8\pi\rho_\infty c_\infty^2}
  \left(\frac{c}{c_\infty}\right)^3\,,
  \label{eq:mag_ram}
\end{equation}
where we have used $r_\mathrm{B}/r_g = c^2/(2c_\infty^2)$.  This ratio
is independent of $M_\mathrm{BH}$ and equals unity when the ambient
field reaches the critical value
\begin{equation}
  B_\mathrm{crit} \sim \sqrt{8\pi\rho_\infty}\,
  \frac{c_\infty^{5/2}}{c^{3/2}}
  \approx 2 \times 10^5\,\mathrm{G}\,.
  \label{eq:B_crit}
\end{equation}
The dominant dynamo-generated field in the radiative solar core is the
Tayler--Spruit dynamo \citep{Spruit2002}, whose poloidal component has a coherence length $L_r \sim 50$\,km $\gg r_\mathrm{B}$ and whose toroidal component vanishes on the rotation axis by axisymmetry.  A PBH at the stellar center therefore sees only the poloidal field, with
$B_r \lesssim 1$\,G (more than five orders of magnitude below
$B_\mathrm{crit}$), giving
$P_\mathrm{mag}/P_\mathrm{ram} \lesssim 10^{-11}$.
Asteroseismic measurements of red giant cores independently constrain
$B_r \lesssim 5$\,kG in the deep interior
\citep{Fuller2015,Stello2016,Cantiello2016,LiGang2022}, still well
below $B_\mathrm{crit}$.

Beyond the instantaneous field strength, one must ask whether flux can accumulate over multiple inflow times. In magnetically arrested disks, the state arises because angular momentum support provides a barrier where flux piles up \citep[e.g.][]{Tchekhovskoy2011}.  In spherical Bondi accretion, no such barrier exists, but the large-scale coherence of the Tayler--Spruit field ($L_r \gg r_\mathrm{B}$) means that field lines are effectively anchored in the nearly stationary stellar plasma at $r \gg r_\mathrm{B}$.  As gas accretes, these field lines are stretched and compressed near the BH but cannot be swallowed whole, allowing flux to build up over time.  Magnetic reconnection, which severs the compressed field lines from their large-scale anchors, acts to limit this accumulation.  Each inflow time $t_\mathrm{in} \sim r_\mathrm{B}/c_\infty$ adds $\sim$$B_0$ worth of flux, giving a constant buildup rate $\dot{B}_\mathrm{build} \sim B_0\,c_\infty/r_\mathrm{B}$. Magnetic reconnection in collisional plasma, as expected in the relevant regime, occurs on a timescale of $t\approx500\,r_g/c$ \citep[e.g.,][]{Bransgrove21}. Therefore, reconnection will destroy the field at a rate $\dot{B}_\mathrm{rec} \sim0.1\,B^2/(r_\mathrm{B}\sqrt{4\pi\rho_\infty})$, which scales as $B^2$ and therefore always overtakes the constant buildup.  Equating the two rates gives an equilibrium field
\begin{equation}
  B_\mathrm{eq} = \sqrt{10\,B_0\,c_\infty\sqrt{4\pi\rho_\infty}}
  \approx 1.5 \times 10^5\,\mathrm{G}\,,
  \label{eq:B_eq}
\end{equation}
for $B_0 \sim 1$\,G, $c_\infty \approx 5 \times 10^7$\,cm\,s$^{-1}$,
and $\rho_\infty \approx 150$\,g\,cm$^{-3}$, corresponding to
$\beta_\mathrm{p,eq} \sim 10^8$, eight orders of magnitude from dynamical importance.

We conclude that for the spherical Bondi accretion considered in this
work, the magnetic field is far too weak to disrupt the flow, both
instantaneously ($P_\mathrm{mag}/P_\mathrm{ram} \lesssim 10^{-11}$)
and even allowing for flux accumulation limited by reconnection
($\beta_\mathrm{p,eq} \sim 10^8$).  The purely hydrodynamic treatment is
therefore well justified throughout the photon-trapping regime and below.
As shown in our companion paper, \citet{Gottlieb2026}, once an accretion disk forms, magnetic processes become relevant, and magnetic fields may play a central role in regulating accretion.

% This might not be needed, but for the time being let's add 
\subsection{Comparison with \citet{Bellinger2023}}
\label{sec:comparison}

\citet{Bellinger2023} studied the evolution of solar-type stars hosting PBHs
using the \texttt{MESA} stellar evolution code, with the PBH accretion
luminosity incorporated as a modified inner boundary condition.  Their
treatment adopts a fixed radiative efficiency $\eta = 0.08$
(the thin-disk value for a Schwarzschild BH) and caps the luminosity at
$L = \min(L_\mathrm{Edd},\,L_\mathrm{B})$, where
$L_\mathrm{B} = \eta \dot{M}_\mathrm{B} c^2$.  Our approach
differs in that we compute $\eta(M_\mathrm{BH})$ self-consistently from
the microphysics of the accretion flow (bremsstrahlung, pair processes,
Klein--Nishina corrections), finding typical values $\eta \approx 0.01$ across most of the mass range, with a local peak $\eta \approx 0.08$ near $10^{-15}\,M_\odot$. The assumed $\eta = 0.08$ is therefore an order of magnitude too high across most of the optically thin range, and in any case is inappropriate for spherical accretion where no disk exists
($r_\mathrm{circ} \ll r_\mathrm{ISCO}$; see also their Eq.~6).

This difference in efficiency has important consequences.  First, the
Eddington limit applies to the \emph{escaping} luminosity only when the
Bondi sphere is optically thick ($\tau \gtrsim 1$), which we show
requires $M_\mathrm{BH} \gtrsim 5 \times 10^{-13}\,M_\odot$;
\citet{Bellinger2023} note that the photon diffusion time is of order
the dynamical time at $r_\mathrm{B}$ (their Eq.~7) but apply
$L < L_\mathrm{Edd}$ as a cap on the accretion rate at all masses.
Second, and more importantly, in the spherical geometry assumed by
\citet{Bellinger2023} (and by us, for $r_\mathrm{circ}\ll r_\mathrm{ISCO}$)
the Eddington limit does \emph{not} cap $\dot{M}$.  As first shown by
\citet{Begelman1979}, the trapping radius
$r_\mathrm{trap} = \dot{M}\kappa/(4\pi c)$ moves outward continuously
from $r_\mathrm{S}$ as $\dot{M}$ rises, so photons produced interior to
$r_\mathrm{trap}$ are advected back to the BH; the escaping luminosity
self-regulates at $L_\mathrm{esc}\lesssim 0.6\,L_\mathrm{Edd}$ but
$\dot{M}\propto M^2$ is preserved (Section~\ref{sec:eddington}).
Consequently, there is no Eddington-limited bottleneck in spherical
accretion: the growth remains super-exponential from
$\tau(r_\mathrm{B})\sim 1$ up to disk formation, and the Salpeter-time
discussion of \citet{Bellinger2023} (their Section~4) does not apply.
Their lower-luminosity model based on the convective Flammang luminosity
$L_\mathrm{F} \sim 10^{-3}\,L_\mathrm{Edd}$ (their Section~4.4) is
qualitatively consistent with the saturated-convection floor we identify
at full trapping (Section~\ref{sec:eddington}), and our microphysical
calculation provides a first-principles justification for this low
effective efficiency.  Overall, our treatment identifies three spherical
regimes (Section~\ref{sec:regimes}) with quantitative transition
criteria, replacing the two-regime ($L_\mathrm{B}$ vs.\ $L_\mathrm{Edd}$)
framework of \citet{Bellinger2023}.

In a companion study, \citet{Caplan2024} extended the Bellinger framework
with an adaptive radiative efficiency that models photon trapping at
high accretion rates, allowing faster BH growth.  This is qualitatively
consistent with our finding that photon trapping reduces the effective
luminosity and preserves the Bondi rate throughout the optically thick
spherical regime, not only above $r_\mathrm{trap}>r_\mathrm{B}$.

Despite these differences, several conclusions are consistent across
studies: the initial Bondi phase dominates the total growth time
($t_\mathrm{B} \propto M_0^{-1}$) and PBHs above a critical initial mass
will inevitably consume their host star.  \citet{Bellinger2023} find
that a PBH more massive than $\sim$$10^{-11}\,M_\odot$ at formation
would alter the observed properties of the present Sun, depending on
the adopted accretion scheme and radiative efficiency.  The critical
mass we find ($M_\mathrm{0,crit} \sim 10^{-16}\,M_\odot$) is
considerably lower, because (i) our self-consistent efficiency
($\eta \sim 10^{-2}$) is an order of magnitude below the assumed
$\eta = 0.08$, and (ii) there is no spherical Eddington
bottleneck, so the partial-trapping window is traversed at the
Bondi rate rather than at the Salpeter rate.  We note that our calculation focuses
on the accretion microphysics and does not include the back-reaction on
the stellar structure, which the \texttt{MESA} calculations of
\citet{Bellinger2023} and \citet{Caplan2024} capture.  A complete
treatment combining our microphysical efficiency with stellar
evolution calculations will be the focus of follow-up work.

\subsection{Relation to other radiating Bondi flow calculations}
\label{sec:radiating_bondi}

The general problem of steady-state spherical accretion with radiative
losses has been studied in several contexts. Early work by
\citet{Shapiro:1973} and \citet{Meszaros:1975} established the basic
framework for the emergent radiation spectrum and the role of
radiation pressure in modifying the inflow. In the
radiation-pressure-dominated, optically thick regime,
\citet{Begelman1978, Begelman:1978} and
\citet{Flammang1982, Flammang:1984} derived analytic and semi-analytic
solutions for accretion onto compact objects at super-Eddington rates,
and \citet{Vitello:1984} performed the first time-dependent numerical
treatment with implicit radiation diffusion. \citet{Park:1990} and
\citet{Nobili:1991} extended these calculations with self-consistent
radiative transfer and, in the case of \citet{Park:1990}, a
two-temperature plasma including thermal pair production
\citep[see also][]{Soffel:1982,Wandel:1984}. More
recently, \citet{2026ApJ...998...22F} revisited the super-Eddington
problem as a model for fallback accretion in failed supernovae,
identifying two distinct analytic solution branches in the
radiation-dominated free-fall limit. In a complementary setting,
\citet{Bailey2026} presented a systematic framework for radiating
Bondi accretion in the gas-pressure-dominated regime, motivated by
giant planet formation in protoplanetary disks.  They solve the same mathematical problem we
address---steady-state spherical accretion modified by radiative
cooling---but in a very different physical regime: protoplanetary gas at
$T \sim 100$\,K with constant dust opacity $\kappa \sim 0.4$\,cm$^2$\,g$^{-1}$,
rather than stellar-interior gas at $T \sim 10^7$\,K with
bremsstrahlung cooling and Klein-Nishina opacity effects.

Despite the different physical context, several structural parallels
are worth noting.  \citet{Bailey2026} parameterize the problem with
a dimensionless cooling time $\beta = t_\mathrm{cool}\,c_\infty/r_\mathrm{B}$,
which plays the same role as our $t_\mathrm{cool}/t_\mathrm{ff}$ ratio in
controlling the transition between the adiabatic and cooling-dominated
regimes.  They identify distinct scaling laws for the accretion
efficiency in the optically thin and optically thick limits, analogous
to our identification of Hot Bondi, bremsstrahlung cooling, and
photon-trapping regimes.  Their ODE shooting method encounters the same
difficulty we find: when cooling is strong, the sonic point structure
changes and steady-state solutions become harder to obtain (in our case,
the sonic point becomes a focus with complex eigenvalues, requiring a
time-dependent approach). This breakdown of smooth steady-state
shooting appears to be a generic feature of strongly cooled spherical
accretion, motivating the time-dependent numerical treatment of
\citet{Vitello:1984} and the boundary-layer analysis of
\citet{2026ApJ...998...22F}.

The key physical differences are threefold.  First, in their
planet-formation context, radiative feedback from the accreting object
suppresses accretion (by heating the surrounding gas), yielding
accretion efficiencies $f_\mathrm{acc} = \dot{M}/\dot{M}_\mathrm{B} < 1$.
In our problem, the PBH has no intrinsic luminosity, and bremsstrahlung
cooling enhances accretion by removing pressure support
($\dot{M}/\dot{M}_\mathrm{B}$ up to $\sim 7$).  Second, their opacities
are constant (dust), while ours vary by orders of magnitude across the
flow due to Klein-Nishina suppression at the relativistic temperatures
($kT \gg m_e c^2$) reached in the inner accretion flow.  Third, our
problem involves a collisionless transition at small radii where the
Coulomb mean free path exceeds the flow scale, a complication absent in
the low-temperature planet-formation context.

\section{Conclusions} \label{sec:conclusions} We have computed the radiative efficiency and accretion rate for spherical accretion onto a primordial black hole embedded in the core of a solar-type star, solving the time-dependent Euler equations with relativistic bremsstrahlung and pair processes across more than six decades in BH mass ($10^{-16.5}$--$10^{-10}\,M_\odot$).  The optically thin Euler solver gives the physical $\eta$ up to $M_\mathrm{BH}\sim 5\times 10^{-13}\,M_\odot$; the higher-mass photon-trapping regime is treated with the analytical prescription of \citet{Begelman1979}. Our main findings are:
\begin{enumerate} 
\item \textbf{Three spherical accretion regimes.} As $M_\mathrm{BH}$ increases, the flow transitions from Hot Bondi (weakly cooled, bremsstrahlung only) through bremsstrahlung-cooled (approaching isothermal) to a single photon-trapping regime in which the locally produced luminosity is largely advected back to the BH \citep{Begelman1979}. There is no spherical Eddington-limited stage: $\dot{M} \approx \dot{M}_\mathrm{B}$ throughout the optically thick range. The transitions are governed by two quantities: the cooling-to-freefall time and the optical depth of the Bondi sphere. A fourth regime, disk accretion, opens once $r_\mathrm{circ}>r_\mathrm{ISCO}$ and is treated in our companion paper \citep{Gottlieb2026}.
\item \textbf{Radiative efficiency $\eta \sim 10^{-2}$.} The self-consistently computed efficiency is typically an order of magnitude below the thin-disk value ($\eta = 0.08$) commonly assumed in the literature. In the Hot Bondi regime, $\eta \approx 0.01$--$0.1$; it decreases to $\eta \approx 0.01$ in the cooling and isothermal limit. The low efficiency reflects the spherical geometry: without a disk, the luminosity is set by bremsstrahlung from the extended flow rather than by dissipation near the ISCO. 
\item \textbf{Enhanced accretion rate.} Bremsstrahlung cooling removes thermal pressure support, increasing the accretion rate by a factor of $\sim$2--7 above the adiabatic Bondi value, saturating at $\dot{M} \approx 7\,\dot{M}_\mathrm{B}$ in the near-isothermal limit. 
\item \textbf{Modest radiative feedback.} Photons deposited outside the Bondi sphere raise the boundary temperature, but convective transport (MLT) limits the enhancement and keeps accretion rates within a factor of $\sim$2 of the unperturbed value. 
\item \textbf{Critical initial mass.} Because the spherical Eddington cap does not apply, growth proceeds at the Bondi rate ($\dot{M}\propto M^2$, super-exponential) throughout the photon-trapping regime. The bottleneck is set entirely by the slow Hot Bondi phase at the lowest masses. The critical initial mass to consume a solar-mass star within a Hubble time is $M_\mathrm{0,crit} \approx 10^{-16}\,M_\odot$, five orders of magnitude below the value implied by previous estimates that assumed thin-disk efficiency and an Eddington cap. \item \textbf{Negligible role of pairs.} Pair production is negligible across the entire mass range, consistent with the two-temperature spherical accretion calculations of \citet{Park:1990}: at low masses the inner flow is collisionless before reaching pair-producing temperatures and cannot maintain thermal pair equilibrium; at higher masses cooling preempts the pair threshold. The luminosity is dominated by bremsstrahlung throughout. \end{enumerate} 
These results have two main implications. First, the low efficiency combined with the absence of a spherical Eddington bottleneck means PBHs grow at the Bondi rate throughout the optically thick range, with $M_\mathrm{0,crit}$ approximately five orders of magnitude below the value found by \citet{Bellinger2023} ($\sim$$10^{-11}\,M_\odot$), suggesting that a broader range of PBH masses may be constrained by observations of solar-type stars. In our companion paper, \citet{Gottlieb2026}, we study the full evolution of the system, including capture, disk formation, and the stellar explosion. Second, our $\eta(M_\mathrm{BH})$ can be incorporated directly into stellar evolution codes as a replacement for the fixed $\eta = 0.08$ assumption. Several directions for future work remain: coupling the accretion model to a stellar evolution code such as \texttt{MESA} \citep{Paxton2011} to capture back-reaction on the stellar structure, and exploring the dependence of $\eta$ on ambient conditions for different stellar types and evolutionary stages.

\section*{Acknowledgments}
  The Center for Computational Astrophysics at the Flatiron Institute is supported by the Simons Foundation. MK and CN are supported by NSF grant PHY-2112839. KVT is supported by the NSF grant PHY-2210551. We thank Mitch Begelman,  Yacine Ali-Haimoud, Andrei Gruzinov, Tamar Faran, Alex Dittmann, Jim Stone, Eliot Quataert, and Andrew MacFadyen for useful discussions.

\appendix

\section{Analytical Fitting Formulae}
\label{sec:analytical_fits}

To facilitate implementation in stellar evolution codes, we provide
analytical fits to the simulation results shown in Figure~\ref{fig:eta}.
The radiative efficiency is well described by a smoothly broken power law
plus a Gaussian component:
\begin{equation}
  \log_{10}\eta = a + \bar{s}\,(x - x_\mathrm{b})
    + \Delta s\,\sqrt{(x - x_\mathrm{b})^2 + \delta^2}
    + A_\mathrm{p}\,e^{-(x - \mu_\mathrm{p})^2/2\sigma_\mathrm{p}^2}\,,
  \label{eq:eta_fit}
\end{equation}
where $x \equiv \log_{10}(M_\mathrm{BH}/M_\odot)$,
$\bar{s} = (s_\mathrm{low} + s_\mathrm{high})/2$, and
$\Delta s = (s_\mathrm{high} - s_\mathrm{low})/2$.  The best-fit
parameters are $a = -1.95$, $s_\mathrm{low} = -0.05$,
$s_\mathrm{high} = 1.18$, $x_\mathrm{b} = -11.76$,
$\delta = 0.19$ for the broken power law, and
$A_\mathrm{p} = 0.61$, $\mu_\mathrm{p} = -15.01$,
$\sigma_\mathrm{p} = 0.60$ for the Gaussian component.
The Gaussian captures a local maximum in $\eta$ near
$10^{-15}\,M_\odot$ that arises from the competition between
accretion rate enhancement and temperature quenching by
bremsstrahlung cooling.  As $M_\mathrm{BH}$ increases through this
range, cooling makes the flow increasingly isothermal at the sonic
point, enhancing $\dot{M}$ by up to a factor $\sim$7 above the
adiabatic Bondi rate (Section~\ref{sec:growth}) and boosting the gas
density throughout the flow.  Since
$\varepsilon_\mathrm{ff} \propto \rho^2$, this density enhancement
increases the luminosity faster than the accretion rate, raising
$\eta$.  At higher masses
($M_\mathrm{BH} \gtrsim 10^{-14.5}\,M_\odot$), cooling quenches the
temperature throughout the flow ($T \to T_\infty$), reducing the
emissivity per unit volume and causing $\eta$ to drop to its minimum
before rising again in the isothermal regime
($\eta \propto M_\mathrm{BH}$).
The overall BPL+Gaussian fit has an RMS residual of 0.10 dex; the
broken-power-law component alone gives 0.31 dex.  We note that the
Gaussian bump tracks a peak in the \emph{with-pairs} simulation data;
the underlying physics suggests pairs are suppressed in the Hot Bondi
regime where this peak sits (Section~\ref{sec:regimes} and
Appendix~\ref{sec:collisionless_pileup}), so the BPL+Gaussian fit
should be regarded as an upper bound on $\eta$ at
$M_\mathrm{BH}\sim 10^{-15}\,M_\odot$, with the BPL-only fit as the
no-pairs lower bound.  In the photon-trapping regime
($M_\mathrm{BH}\gtrsim 5\times 10^{-13}\,M_\odot$), the fitted $\eta$
gives the locally emitted efficiency in the optically thin Bondi+cooling
limit; the externally observed efficiency is capped by the
\citet{Begelman1979} bound $L_\mathrm{esc}\lesssim 0.6\,L_\mathrm{Edd}$
(Section~\ref{sec:eddington}).

For implementation in stellar evolution codes, the recommended
procedure (equivalent to Equation~\ref{eq:Mdot_actual}) is:
\begin{enumerate}
\item Compute the adiabatic Bondi rate $\dot{M}_\mathrm{B}$ and the
  cooling-enhanced isothermal-limit rate $f(M)\,\dot{M}_\mathrm{B}$
  using Equation~\ref{eq:f_fit}.  Compute the locally emitted
  luminosity $L_\mathrm{emitted} = \eta(M)\,f(M)\,\dot{M}_\mathrm{B}\,c^2$
  using Equation~\ref{eq:eta_fit}.
\item Compute the trapping radius
  $r_\mathrm{trap} = \dot{M}_\mathrm{B}\,\kappa/(4\pi c)$ and the
  geometric escape fraction
  $f_\mathrm{geom} = \max(0,\,1-r_\mathrm{trap}/r_\mathrm{B})$.
  Compute the escaping luminosity
  $L_\mathrm{esc} = f_\mathrm{geom}\,\min(L_\mathrm{emitted},
  0.6\,L_\mathrm{Edd})$ and the cooling efficiency
  $\eta_\mathrm{cool} = L_\mathrm{esc}/L_\mathrm{emitted}$.
\item Set
  $\dot{M} = \dot{M}_\mathrm{B}\,
            \bigl[1 + (f(M)-1)\,\eta_\mathrm{cool}\bigr]\,
            (1 - L_\mathrm{esc}/L_\mathrm{Edd})^2$.
\item If a disk forms ($r_\mathrm{circ}>r_\mathrm{ISCO}$), switch to
  the standard thin-disk Eddington-limited rate
  $\dot{M}_\mathrm{Edd}^\mathrm{disk} = L_\mathrm{Edd}/(\eta_\mathrm{disk}c^2)$
  with $\eta_\mathrm{disk}\approx 0.08$ \citep[treated in our
  companion paper,][]{Gottlieb2026}.
\end{enumerate}

The accretion rate enhancement $f(M) \equiv \dot{M}/\dot{M}_\mathrm{B}$
from radiative cooling is fit by a log-Gaussian:
\begin{equation}
  \log_{10} f = c + A\,\exp\!\left[
    -\frac{(x - \mu)^2}{2\sigma^2}\right]\,,
  \label{eq:f_fit}
\end{equation}
with $c = -0.05$, $A = 0.85$, $\mu = -14.68$, and $\sigma = 0.83$.
At low masses ($M_\mathrm{BH} \lesssim 10^{-16}\,M_\odot$),
$f \approx 1.6$; the enhancement rises through the transition region
and saturates at $f \approx 7$ for
$M_\mathrm{BH} \gtrsim 10^{-14}\,M_\odot$ where the flow is fully
isothermal.  Note that the log-Gaussian captures the rising branch
but underestimates $f$ at high masses where the enhancement
plateaus; for $M_\mathrm{BH} \gtrsim 10^{-14}\,M_\odot$, $f \approx 7$
should be used directly.  This fit applies in the optically thin
regime; in the photon-trapping regime (Section~\ref{sec:regimes}),
$\dot{M}$ is set self-consistently via Equation~\ref{eq:Mdot_actual},
which smoothly reduces the cooling enhancement as photons are advected
inward.

\section{Collisionless Inner Region: Loss Cone, Recycling, and Self-Induced Collisionalization}
\label{sec:collisionless_pileup}

The lightest masses in our range
($M_\mathrm{BH} \lesssim 10^{-14}\,M_\odot$) sit in the Hot Bondi
regime of Section~\ref{sec:regimes}: the gas is collisional at the
Bondi radius, but the Coulomb mean free path becomes comparable to the
flow scale at a transition radius
$r_\mathrm{coll}$ well inside~$r_\mathrm{B}$. For the fiducial case
$M_\mathrm{BH} = 10^{-16}\,M_\odot$ in the solar core, with
$T_\infty = 1.57\times 10^{7}\,\mathrm{K}$ and
$\rho_\infty = 150\,\mathrm{g\,cm^{-3}}$, we find
$r_\mathrm{coll} \simeq 7000\,r_\mathrm{S}$, still subsonic (Mach
number $\approx 0.7$) and well outside the cooling-induced sonic
point at $r_\mathrm{sonic} \simeq 600$--$800\,r_\mathrm{S}$
(Figure~\ref{fig:schematic_pileup}).

Inside $r_\mathrm{coll}$ the fluid approximation is no longer valid, and individual particles retain their thermal transverse velocities. Most carry specific angular momentum well above the GR capture threshold
$\ell_\mathrm{crit} = 4 G M_\mathrm{BH}/c$, miss the loss cone on any
single pass, and would be lost from the accretion supply if the
problem were fully collisionless. We show below that this is not
what happens: the missed particles recycle and self-collisionalize, and
the only effect on the hydrodynamic supply is a moderate pressure
feedback localized to the lightest masses.

\subsection{Single-pass capture and recycling}
\label{sec:capture_recycling}

\begin{figure}[t]
  \centering
  \includegraphics[width=\columnwidth]{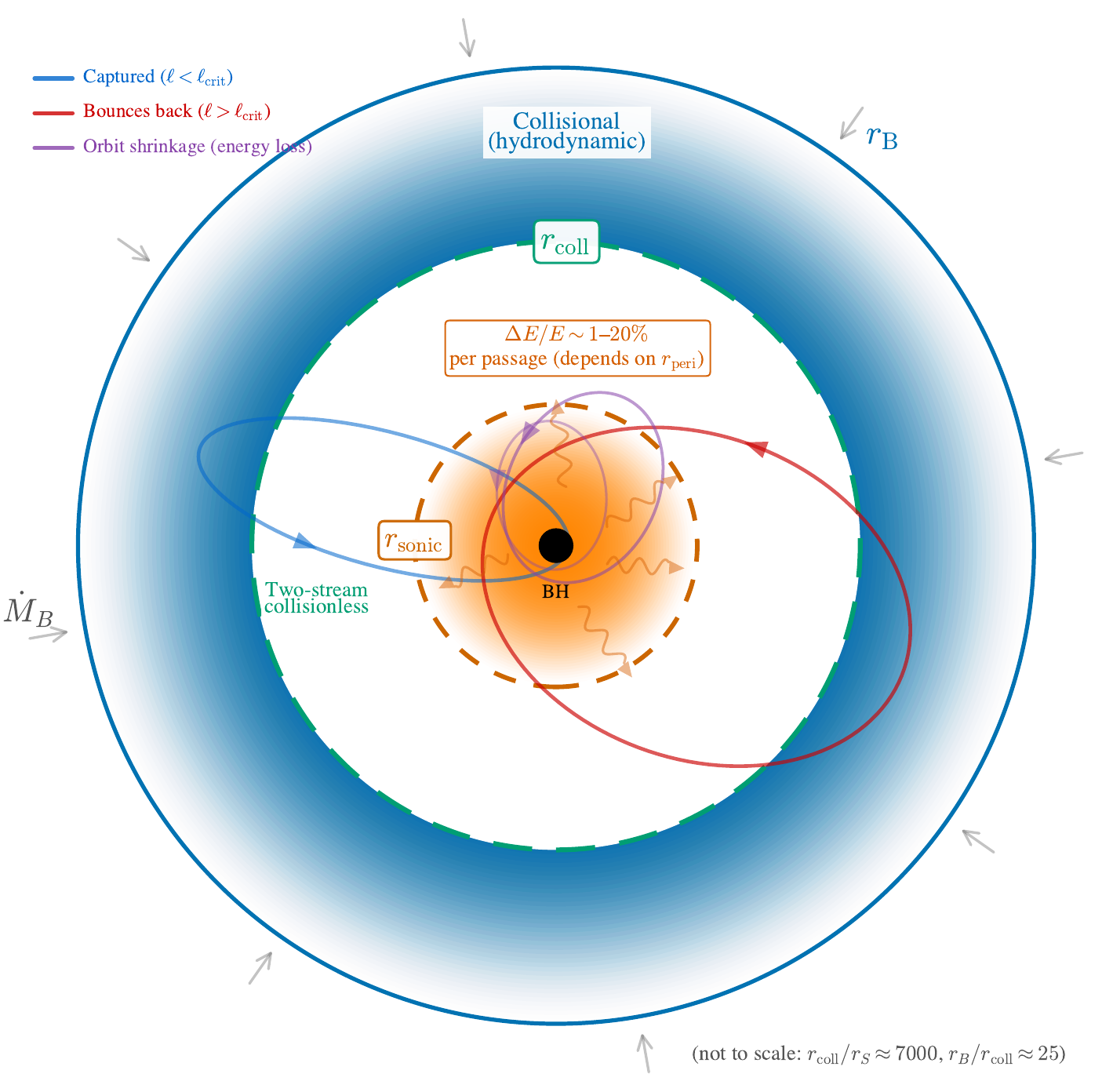}
  \caption{Schematic of the recycling/pile-up geometry inside the
  collisionless region. The flow is hydrodynamic near $r_\mathrm{B}$
  and becomes collisionless below $r_\mathrm{coll}$. Most particles
  miss the GR loss cone on a single inward pass, return to
  $r_\mathrm{coll}$, re-thermalize, and either are recaptured or
  circularize after losing orbital energy to bremsstrahlung. Direct
  GR capture (loss cone) accounts for $\sim 0.2\%$ of entrants;
  $\sim 33\%$ are reflected at the centrifugal barrier, and $\sim 67\%$
  enter the recycling/circularization population.}
  \label{fig:schematic_pileup}
\end{figure}

At $r_\mathrm{coll}$ the bulk flow is radial but particles retain a
one-component transverse velocity dispersion
$\sigma_\perp = c_s/\sqrt{\gamma}$, giving a typical specific angular
momentum
$\ell_\mathrm{typ} \sim r_\mathrm{coll}\,\sigma_\perp
\simeq 30\,\mathrm{cm^{2}\,s^{-1}}$, while
$\ell_\mathrm{crit} \simeq 1.8\,\mathrm{cm^{2}\,s^{-1}}$ for
$M_\mathrm{BH} = 10^{-16}\,M_\odot$. The Rayleigh distribution of
$|\ell|$ then implies a single-pass loss-cone fraction
\begin{equation}
  f_\mathrm{cap} \simeq \tfrac{1}{2}
  (\ell_\mathrm{crit}/\ell_\mathrm{typ})^2
  \simeq 1.5\times 10^{-3}.
  \label{eq:fcap}
\end{equation}
Taken at face value this small number would suppress
$\dot{M}/\dot{M}_\mathrm{B}$ by the fully collisionless factor
$(c_s/c)^2 \sim 3\times 10^{-6}$. It does not, because missed
particles do not escape: they swing back through $r_\mathrm{coll}$,
re-enter the collisional region, re-thermalize, and are re-supplied to
the inflow on subsequent crossings (Figure~\ref{fig:schematic_pileup}).
In a steady state with effective per-crossing capture/transport
fraction $f_\mathrm{eff}$, the inward flux at $r_\mathrm{coll}$ is
$\dot{M}/f_\mathrm{eff}$ and the residence-time density enhancement is
$\xi \simeq 2/f_\mathrm{eff}$.

\subsection{Bremsstrahlung self-limits the pile-up}
\label{sec:brems_xi}

Orbits inside $r_\mathrm{coll}$ are not conservative.
Bremsstrahlung radiates a fraction
\begin{equation}
  (\Delta E/E)_0 \simeq
  \frac{\varepsilon_\mathrm{ff}}{\rho}\,\frac{r}{v}\,\frac{1}{\tfrac{1}{2}v^2}
\end{equation}
per pericenter passage at the unperturbed density. Integrating along
full Newtonian orbits in our converged cooled background, and weighting
by the angular-momentum distribution of the recycling population,
gives
\begin{equation}
  \langle\Delta E/E\rangle_\mathrm{rec} \simeq 0.018,
  \qquad
  \langle\xi\rangle_\mathrm{rec} \simeq 12.
\end{equation}
The pile-up is self-limited because at enhanced density $\xi\rho$ the
cooling rate per unit mass scales as $\xi$, so the residence time
shrinks and a higher pile-up cools itself faster. Combining
$\xi \simeq 2 N$ with $N \simeq [\xi(\Delta E/E)_0]^{-1}$ gives
$\xi \simeq \sqrt{2/(\Delta E/E)_0} \sim 5$--$13$.

This energy loss does not in itself drive particles into the loss
cone: bremsstrahlung removes orbital energy more efficiently than
angular momentum, so missed particles instead circularize at
$r_\mathrm{circ}(\ell) \simeq \ell^2/(GM_\mathrm{BH})$. Combined with
the small per-pass capture probability~(\ref{eq:fcap}) and the typical
$N \sim 6$ orbital passages before circularization, the cumulative
probability of direct GR capture before circularization is only of
order $1\%$. Direct capture is therefore not the dominant sink: the
key question is whether the circularized material remains
collisionless, which would build a permanent reservoir of trapped
particles, or self-collisionalizes.

\subsection{Self-induced collisionalization}
\label{sec:self_coll}

\begin{figure}[t]
  \centering
  \includegraphics[width=\columnwidth]{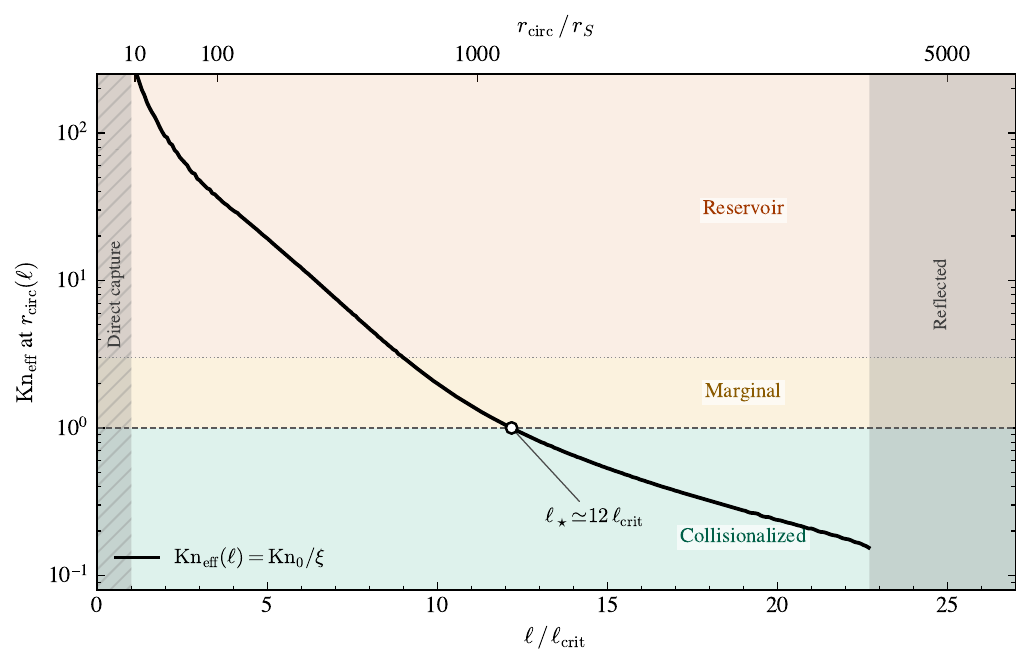}
  \caption{Effective Knudsen number at the circularization radius as a
  function of specific angular momentum, for
  $M_\mathrm{BH} = 10^{-16}\,M_\odot$. The hatched bands mark the
  excluded loss-cone ($\ell < \ell_\mathrm{crit}$) and reflected
  ($\ell > \ell_\mathrm{max}$) populations. The pile-up enhancement
  $\xi(\ell)$ is sufficient to make
  $\mathrm{Kn}_\mathrm{eff} < 1$ for the bulk of the
  recycling population ($\ell \gtrsim \ell_\star \simeq
  12\,\ell_\mathrm{crit}$), so most circularization radii are
  self-collisionalized. Only a low-$\ell$ tail
  ($\sim 21\%$) remains in the formally collisionless ``reservoir''
  regime; the additional mass needed to push these bins to
  $\mathrm{Kn}_\mathrm{eff} \sim 1$ is
  $\sim 6\times 10^{-15}\,\mathrm{g}$ and accumulates in only a few
  free-fall times.}
  \label{fig:knmap}
\end{figure}

The circularized material increases the local density. Since the
Coulomb mean free path scales approximately as
$\lambda_\mathrm{mfp}\propto \rho^{-1}$, the pile-up lowers the
effective Knudsen number at the circularization radius,
\begin{equation}
  \mathrm{Kn}_\mathrm{eff}(\ell) =
    \mathrm{Kn}_0[r_\mathrm{circ}(\ell)] / \xi(\ell),
\end{equation}
where $\mathrm{Kn}_0 = \lambda_\mathrm{mfp}/r$ is computed from the
unperturbed cooled profile. Figure~\ref{fig:knmap} shows the result
for $M_\mathrm{BH} = 10^{-16}\,M_\odot$.

The result is a self-regulating sequence: missed particles cool and
circularize, the resulting pile-up raises the density, the higher
density shortens the Coulomb mean free path, and the inner region
becomes collisional enough to redistribute angular momentum and
support hydrodynamic transport. The recycling loop closes; no
permanent collisionless reservoir survives.

\subsection{Hydrodynamic back-reaction and the corrected
$\dot{M}$}
\label{sec:back_reaction}

Although the recycling loop closes, the pile-up at $r_\mathrm{coll}$
exerts a back-pressure on the surrounding flow. Because
$r_\mathrm{coll} > r_\mathrm{sonic}$ for the lowest masses, this
pressure perturbation lies in the subsonic region and can communicate
upstream, modifying the Bondi eigenvalue. We model this by replacing
the kinetic region with conservative momentum and energy source
terms~$S_p(r)$, $S_E(r)$ derived from the orbit-averaged kinetic
closure, and inject them into the one-dimensional Euler solver. The
closure is updated periodically from the evolving hydrodynamic state
with under-relaxation ($\omega = 10^{-3}$, recomputed every
$10^3$ hydro steps), allowing the kinetic feedback to follow the
secular adjustment of the flow without driving spurious oscillations.

The system converges monotonically to a self-consistent fixed point.
Table~\ref{tab:chi_M} summarizes the resulting accretion-rate
correction
$\chi(M_\mathrm{BH}) \equiv \dot{M}/\dot{M}_\mathrm{B}$ across the
lightest masses.

\begin{table}[h]
\centering
\caption{Self-consistent accretion-rate correction
  $\chi \equiv \dot{M}/\dot{M}_\mathrm{B}$ from the hydro-kinetic
  source coupling, and the radial ordering
  $r_\mathrm{coll}/r_\mathrm{sonic}$ that controls whether the
  pile-up sits in the subsonic zone.}
\label{tab:chi_M}
\begin{tabular}{cccc}
\hline\hline
$\log_{10}(M_\mathrm{BH}/M_\odot)$ & $\chi$ & reduction &
$r_\mathrm{coll}/r_\mathrm{sonic}$ \\
\hline
$-16.0$ & 0.69 & $31\%$        & $8.9$ \\
$-15.8$ & 0.95 & $5\%$         & $11.5$ \\
$-15.6$ & 0.94 & $6\%$ (marg.) & $0.94$ \\
$-15.3$ & 1.00 & $0\%$         & $0.06$ \\
\hline
\end{tabular}
\end{table}

The correction is sharply localized to the lowest masses. Above
$M_\mathrm{BH} \simeq 10^{-15.5}\,M_\odot$, the cooling-induced sonic
point moves outside $r_\mathrm{coll}$, the pile-up sits in the
supersonic zone where it cannot communicate upstream, and the
one-dimensional Bondi+cooling calculation is fully self-consistent
($\chi = 1$). The growth integration in
Section~\ref{sec:growth} therefore uses the standard $\dot{M}_\mathrm{B}$
across the entire range; the only mass at which $\chi$ departs from
unity at the level of tens of percent is the lightest one
($M_\mathrm{BH} \sim 10^{-16}\,M_\odot$). Because $\chi(M)$ recovers
to unity over a narrow window in mass ($\Delta\log M\!\approx\!0.5$),
while the Bondi growth time $t_\mathrm{B}\propto M^{-1}$ accumulates
predominantly at the lightest end of the trajectory, applying the
largest correction ($\chi\!\simeq\!0.7$) to the leg at
$10^{-16}\,M_\odot$ shifts $M_\mathrm{0,crit}$ by only $\sim$30\% in
mass ($\sim$0.1 dex)---well within the uncertainty introduced by
ambient stellar conditions.

\subsection{Anomalous collisionality from plasma
microinstabilities}
\label{sec:anomalous}

The kinetic closure above includes only Coulomb scattering. The
recycling region, however, is intrinsically counter-streaming: fresh
inflowing material, outbound bouncers, and circularizing particles
coexist in the same volume with substantial relative velocities. Such
configurations are well known to be unstable to current-driven and
Weibel-type plasma microinstabilities \citep{Weibel:1959}, which
generate magnetic field \emph{de novo} from noise on the electron skin
depth $\delta_e = c/\omega_p$. For the fiducial
$10^{-16}\,M_\odot$ case at solar-core densities,
$\delta_e \sim 5\times 10^{-8}\,\mathrm{cm}$, smaller than
$r_\mathrm{coll} \sim 2\times 10^{-7}\,\mathrm{cm}$. Equipartition
with the bulk kinetic energy would correspond to
$B_\mathrm{eq} \sim 4\times 10^{8}\,\mathrm{G}$; even at saturation
levels well below equipartition ($\epsilon_B \sim 10^{-6}$), the
proton Larmor radius is $r_L \sim 10^{-9}\,\mathrm{cm}$, deeply
sub-$r_\mathrm{coll}$. The plasma is then magnetized on every scale
that matters, and pitch-angle scattering off the wave field provides
anomalous collisionality, much as in collisionless shocks.

Anomalous scattering can only \emph{add} to the Coulomb collisionality
already accounted for in Section~\ref{sec:self_coll}, and its effects
all push $\chi$ towards unity: the low-$\ell$ ``reservoir'' bin in
Figure~\ref{fig:knmap} collisionalizes faster than free-fall, part of
the marginal $\mathrm{Kn}_\mathrm{eff} \sim 1$--$3$ population is
moved into the collisional regime, and angular-momentum redistribution
near $r_\mathrm{coll}$ becomes more efficient. The values of $\chi$ in
Table~\ref{tab:chi_M} should therefore be read as a conservative lower
bound: pure Coulomb collisions give the strongest pile-up
back-pressure, while any plasma-microinstability contribution moves
the system closer to fully hydrodynamic accretion at
$\dot{M}_\mathrm{B}$. A quantitative determination of the saturated
$\epsilon_B$ and of the resulting pitch-angle scattering rate would
require a dedicated kinetic (PIC-level) calculation and is beyond the
scope of this paper.

\subsection{Summary}

For the masses considered in the main text, the gas is collisional at
$r_\mathrm{B}$, so a hydrodynamic Bondi supply is established. The
flow becomes collisionless only at small radii
($r_\mathrm{coll} \sim 7000\,r_\mathrm{S}$ at
$10^{-16}\,M_\odot$), where most particles miss the GR loss cone on a
single pass. Bremsstrahlung-driven circularization, the resulting
pile-up, and self-induced collisionalization keep the inner region
from acting as a permanent reservoir; at the same time, the pile-up
exerts a moderate back-pressure on the subsonic Bondi flow that
reduces $\dot{M}/\dot{M}_\mathrm{B}$ by at most $\sim 30\%$ at the
lightest mass and by less than a few percent for
$M_\mathrm{BH} \gtrsim 10^{-15.8}\,M_\odot$. Plasma
microinstabilities in the counter-streaming inner flow are expected to
add anomalous collisionality on top of the Coulomb estimate, so the
correction quoted here is conservative. The fully collisionless
$(c_s/c)^2$ suppression applies in principle once $r_\mathrm{B}$
itself is collisionless, at
$M_\mathrm{BH} \lesssim 10^{-18}\,M_\odot$, well below our range; even
there, anomalous collisionality from the same plasma microinstabilities
discussed in Section~\ref{sec:anomalous} is expected to soften this
scaling, so $(c_s/c)^2$ should likewise be read as a lower bound on
$\dot{M}/\dot{M}_\mathrm{B}$.

%----------------------------------------------------------------------
\section{Simulation Parameters and Results}
\label{sec:sim_table}

Table~\ref{tab:simulations} summarizes the numerical setup and key
outputs for each black hole mass computed in this work.  At each mass,
the radiative efficiency $\eta$ and accretion rate enhancement $f$ are
taken from the best-available feedback model: the diffusion model
(Section~\ref{sec:radiative_feedback}) where the feedback parameter
$\beta < 1$, and the MLT convective envelope model where $\beta \geq 1$.
The peak temperature $T_\mathrm{max}$ is measured from the baseline
(no-feedback) simulation, which sets the microphysical emissivity
profile.  We tabulate only the optically thin masses
($M_\mathrm{BH} \lesssim 5\times 10^{-13}\,M_\odot$), for which the
simulation $\eta$ is the physical radiative efficiency.  At higher
masses the Bondi sphere becomes optically thick and photon trapping
caps $L_\mathrm{esc}\lesssim 0.6\,L_\mathrm{Edd}$ (Section~\ref{sec:eddington});
in that regime the optically thin solver no longer provides a
self-consistent $\eta$, and the growth calculation instead uses the
analytical prescription of Equation~\eqref{eq:Mdot_actual}.

\begin{table*}
\centering
\caption{Simulation parameters and results for each PBH mass.
  Columns: (1)~black hole mass; (2)~radiative efficiency from the
  best-available feedback model; (3)~accretion rate enhancement over
  the adiabatic Bondi rate; (4)~peak gas temperature in units of the
  ambient temperature $T_\infty = 1.57 \times 10^7$\,K;
  (5)~luminosity; (6)~feedback parameter $\beta$
  (Equation~\ref{eq:beta_feedback}); (7)~feedback model used
  (Diff = diffusion, MLT = mixing length theory);
  (8)~number of radial grid cells; (9)~inner boundary radius in units
  of $r_\mathrm{B}$.}
\label{tab:simulations}
\begin{tabular}{ccccccccc}
\hline\hline
$\log_{10}\!\left(\frac{M_\mathrm{BH}}{M_\odot}\right)$ &
$\eta$ &
$f$ &
$T_\mathrm{max}/T_\infty$ &
$L$\,[erg\,s$^{-1}$] &
$\beta$ &
Model &
$N$ &
$r_\mathrm{in}/r_\mathrm{B}$ \\
\hline
$-16.1$ & $2.44\times10^{-2}$ & 1.47 & $2.5\times10^{4}$ & $9.0\times10^{18}$ & $4.5\times10^{-5}$ & Diff & 800 & $3\times10^{-6}$ \\
$-16.0$ & $2.66\times10^{-2}$ & 1.63 & $2.6\times10^{4}$ & $1.5\times10^{19}$ & $7.7\times10^{-5}$ & Diff & 800 & $3\times10^{-6}$ \\
$-15.6$ & $4.18\times10^{-2}$ & 2.74 & $1.9\times10^{4}$ & $1.5\times10^{20}$ & $7.6\times10^{-4}$ & Diff & 800 & $3\times10^{-6}$ \\
$-15.3$ & $6.13\times10^{-2}$ & 4.00 & $1.4\times10^{4}$ & $9.0\times10^{20}$ & $4.5\times10^{-3}$ & Diff & 800 & $3\times10^{-6}$ \\
$-15.1$ & $7.58\times10^{-2}$ & 4.79 & $1.2\times10^{4}$ & $2.8\times10^{21}$ & $1.4\times10^{-2}$ & Diff & 800 & $3\times10^{-6}$ \\
$-15.0$ & $8.30\times10^{-2}$ & 5.16 & $1.0\times10^{4}$ & $4.8\times10^{21}$ & $2.4\times10^{-2}$ & Diff & 800 & $3\times10^{-6}$ \\
$-14.5$ & $3.81\times10^{-2}$ & 6.22 & $2.1\times10^{3}$ & $2.0\times10^{22}$ & $1.0\times10^{-1}$ & Diff & 6400 & $5\times10^{-6}$ \\
$-14.3$ & $1.94\times10^{-2}$ & 6.26 & $9.3\times10^{2}$ & $2.8\times10^{22}$ & $1.4\times10^{-1}$ & Diff & 6400 & $5\times10^{-6}$ \\
$-14.0$ & $4.34\times10^{-2}$ & 7.18 & $2.0\times10^{2}$ & $1.2\times10^{23}$ & $8.2$ & MLT & 6400 & $5\times10^{-6}$ \\
$-13.5$ & $1.75\times10^{-2}$ & 7.16 & $26$ & $2.7\times10^{23}$ & $1.3\times10^{2}$ & MLT & 6400 & $10^{-5}$ \\
$-13.3$ & $1.48\times10^{-2}$ & 7.15 & $11$ & $4.6\times10^{23}$ & $8.3\times10^{2}$ & MLT & 6400 & $10^{-5}$ \\
$-13.0$ & $1.20\times10^{-2}$ & 7.13 & $5$ & $1.1\times10^{24}$ & $7.3\times10^{3}$ & MLT & 6400 & $10^{-5}$ \\
$-12.5$ & $1.05\times10^{-2}$ & 7.10 & $14$ & $5.6\times10^{24}$ & $2.3\times10^{4}$ & MLT & 800 & $10^{-5}$ \\
\hline
\end{tabular}
\end{table*}

\bibliographystyle{aasjournal}
\bibliography{references}

\end{document}